\documentclass[%
 aip,
 amsmath,amssymb,
 reprint,%
]{revtex4-1}

\usepackage{graphicx}
\usepackage{dcolumn}
\usepackage{bm}

\usepackage[utf8]{inputenc}
\usepackage[T1]{fontenc}
\usepackage{mathptmx}

\usepackage[normalem]{ulem} 

\begin{document}

	\author{P.I. Kos}
	\email{kos@polly.phys.msu.ru}
	\affiliation{Faculty of Physics, Lomonosov Moscow State University, 119991 Moscow, Russia}
	\affiliation{Semenov Institute of Chemical Physics RAS, 119991, Moscow, Russia}

	\author{V.A. Ivanov}
	\affiliation{Faculty of Physics, Lomonosov Moscow State University, 119991 Moscow, Russia}
	\affiliation{Institute of Physics, Martin Luther University, 06099 Halle (Saale), Germany}
	
	\author{A.V. Chertovich}
	\affiliation{Faculty of Physics, Lomonosov Moscow State University, 119991 Moscow, Russia}
	\affiliation{Semenov Institute of Chemical Physics RAS, 119991, Moscow, Russia}

\title[Crystallization in melts and poor-solvent solutions of semiflexible polymers: extensive DPD study]{Crystallization in melts and poor-solvent solutions of semiflexible polymers: extensive DPD study}



\date{\today}

\begin{abstract}
In the present work, the process of crystallization of semiflexible polymer chains in melts and poor-solvent solutions with different concentration was studied using dissipative particle dynamics (DPD) simulation technique. We used the coarse-grained polymer model to reveal the general principles of crystallization in such systems at large scales of time and length. It covers both primary and secondary nucleation as well as crystallites' merging. The parameters of the DPD model were chosen appropriately to reproduce the entanglements of polymer chains. We started from an initial homogeneous disordered polymer solution and observed the crystallization process without and with polymer-solvent separation. We have found that the overall crystalline fraction at the end of crystallization process decreases with increasing of the polymer volume fraction. The steady-state crystallization speed at later stages is almost the same for all of the polymer volume fractions. The average crystallite size has a maximal value in the systems with polymer volume fraction from 70\% to 95\% . In our model, these polymer concentrations represent an optimal value in the sense of balance between the amount of polymer material available to increase crystallite size and chain entanglements, that prevent crystallites' growth and merging. We observed a logarithmic in time lamellar thickening.
\end{abstract}

\maketitle

	\section{\label{intro}INTRODUCTION}
	
	Crystallization of polymer materials influences strongly their macroscopic properties and plays an important role in many applications~\cite{mandelkern2002crystallization, mandelkern2004crystallization, SperlingBook, StroblBook, ReiterStrobl, muthukumar2004review, Crystals2017}. Understanding of polymer crystallization on the molecular level is one of the most challenging unsolved problem in modern polymer physics. Nowadays the detailed knowledge of polymer crystallization mechanism still stays unclear for many semi-crystalline polymers~\cite{mandelkern2002crystallization, mandelkern2004crystallization, SperlingBook, StroblBook, ReiterStrobl, muthukumar2004review, Crystals2017}. From the microscopic point of view the main difference between polymer crystallization and low-molecular inorganic crystallization is the fact that monomers are constrained by connecting in chains, so that the concept of a polymer chain conformation and ``chain folding''~\cite{mandelkern2002crystallization, mandelkern2004crystallization, SperlingBook, StroblBook, ReiterStrobl, muthukumar2004review, Crystals2017} becomes important. In the melt of long non-phantom chains a network of entanglements is formed, and this fact dramatically slows down the dynamics and is responsible for many important mechanical features such as, e.g., high elasticity of polymer materials~\cite{DoiEdwards}. In addition, this network of entanglements makes impossible the complete crystallization in systems with long polymer chains, so that polymers which are able to crystallize always stay semi-crystalline~\cite{mandelkern2002crystallization, mandelkern2004crystallization, SperlingBook, StroblBook, ReiterStrobl, muthukumar2004review, Crystals2017}.
	
	Polymer crystallization is a very complex phenomenon, and it includes many different aspects~\cite{mandelkern2002crystallization, mandelkern2004crystallization, SperlingBook, StroblBook, ReiterStrobl, muthukumar2004review, Crystals2017}. One should distinguish crystallization from solutions and from melts, primary and secondary nucleation processes during crystallization, homogeneous vs. heterogeneous nucleation, consider possible contributions from equilibrium thermodynamics and from kinetics, evidences for nucleation and growth scenario and spinodal decomposition scenario at different time scales, possible presence of ``precursors'' (preordered mesomorphic structures and may be even a mesomorphic phase), take into account chain length, value of supercooling, etc.~\cite{mandelkern2002crystallization, mandelkern2004crystallization, SperlingBook, StroblBook, ReiterStrobl, muthukumar2004review, Crystals2017} For such a complex phenomenon, it is not surprisingly, that there are controversies in theoretical and computer simulation literature on some of these aspects, so that we seem to be still quite far from complete understanding of polymer crystallization in different systems~\cite{Crystals2017}.
	
	There is still no generally accepted analytical description of polymer crystallization based on the microscopic approach of statistical physics. Crystallization is a 1st-order phase transition, so that there is no universality in the sense of universality in critical phenomena. It is hard to expect that a unified description of crystallization process can be found for all crystallizing polymers. However, there are universal properties of all polymer systems -- connectivity of monomer units in chains, intrachain stiffness, topology (entanglements) -- and therefore one could expect some general features (i.e., some similarity) in crystallization behavior of particular classes of polymers. Such general features could be, e.g., a particular scenario of crystallization, initial spontaneous thickness of a lamella, its dependence on temperature, lamellar thickening with time, etc.~\cite{mandelkern2002crystallization, mandelkern2004crystallization, SperlingBook, StroblBook, ReiterStrobl, muthukumar2004review, Crystals2017}
	
	 In general, most of theoretical approaches, that describe polymer crystallization, do not correspond to some microscopic view of polymer crystallization at the level of chain conformation and its statistical behavior~\cite{Crystals2017}. Several successful attempts to describe different aspects of crystallization in polymers by means of statistical physics approach are known for a long time~\cite{Allegra1977, Muthukumar2003}, including also studies of chain statistics in polymer crystallization~\cite{Allegra2009}. Recently, a simple kinetic model of a polymer crystalline lamella formation has been proposed~\cite{Stepanow2014}, based on competition between coil and rod-like conformations in overcooled polymer melt, and this kinetic theory predicts correct lamella thickness depending on temperature, similar to more phenomenological approach by Strobl~\cite{Strobl2000, Strobl2006, Strobl2009}.

	 Let us mention here just a few open problems in understanding polymer crystallization. For a process from solution, the dependence of mechanisms and properties on molecular weight and on polymer concentration is still partially understood and various explanations coexist~\cite{Crystals2017, KundMuthu2007}. For early stages of polymer crystallization (both from solutions and from melts), the nucleation and growth (NG) scenario is generally considered as being realized in most polymer systems, and some mesomorphic preordered structures, like ``baby nuclei''~\cite{Muthukumar2003}, bundles~\cite{Allegra1999}, precursor layers~\cite{Luo2011} are formed. The mechanisms of growing and possibly also merging of such nuclei are still poorly understood~\cite{Crystals2017}. There is even the concept of a mesomorphic phase~\cite{Strobl2000, Strobl2006, Strobl2009}, which, however, has been criticized~\cite{Muthukumar2000,Luo2011}. Moreover, there is still a confrontation in the literature regarding the dominance of various scenarios, i.e., nucleation and growth (NG) versus spinodal decomposition (SD), at early stages of primary nucleation~\cite{Crystals2017,muthukumar2004review,gee2006atomistic}.
	
	Computer simulations which deal with microscopic models can shed light on many aspects of polymer crystallization. Microscopic models can be atomistic or coarse-grained (CG) depending on which particular aspects of behavior of a real system one would like to investigate. Advantages and disadvantages of atomistic and CG approaches are well known, and nowadays multiscale simulations are required for solution of most problems~\cite{Elliott2011,Gooneie-Polymers2017}. Atomistic and ``united atom'' (UA) models are suitable to reveal microscopic mechanisms of crystallization in a particular polymer systems, however, such models have been constructed so far for only a few polymers, e.g., for polyethylene (PE), polyvinyl alcohol (PVA) and poly(vinylidene fluoride) (pVDF)~\cite{paul1995optimized, Yamamoto1997, Yamamoto1998I, Rutledge2002, Rutledge2004, Meyer2001, Meyer2002, vettorel2006coarse, Vettorel2007, gee2006atomistic}. More rough or ``larger-grain'' CG models (e.g., bead-spring model where one bead includes several monomeric units of a polymer chain) intend to reveal general features (i.e., those depending only on universal properties of polymers) of crystallization in a broad class of polymer systems, e.g., in semiflexible polymers, or in comb-like polymers, etc. The usual way to induce polymer crystallization in different CG models is to increase the chain stiffness by using torsion and/or bond angle potentials, which stimulate chains being packed into lamellae. CG models can be good enough to capture some properties of real polymers, but they can fail in the description of their other important features (a well known example is the stability of rotator phase in UA model of PE~\cite{paul1995optimized}, while the orthorhombic phase cannot be obtained using this model~\cite{paul1995optimized, Paul-Shakirov}). Computer simulations of polymer crystallization using different models were performed by several groups using both molecular dynamics (MD) and Monte Carlo (MC) methods~\cite{Meyer2001, Meyer2002, vettorel2006coarse, Muthukumar1998, Welch2001, muthukumar2005modeling,Zhang2007, welch2017examining, Yamamoto1997, Yamamoto1998I, Yamamoto2008, Yamamoto2009, Yamamoto2010, Yamamoto2013, Rutledge2002, Rutledge2004, Rutledge2007a, Yi2013, yeh2015mechanical, bourque2016molecular, Vettorel2007, Sommer2007a,Luo2014, luo2016entanglements, anwar2013crystallization, anwar2015crystallization, Paul-Shakirov, Wang2015}.
    
    General theoretical concept of folded chains in lamellae crystallized both from solutions and from melts is now well established~\cite{Crystals2017}, and this concept has been many times (actually, since the discovery of crystallization in polymers more than 60 years ago) confirmed in experimental studies~\cite{Sadler77, Fischer1984, Fischer1988, HongPRL2015, HongMacromol2015, HongACSLett2016}. In computer simulations, chain folding in the course of polymer crystallization has been studied both in melts~\cite{Yamamoto1997, Yamamoto2008, Yamamoto2009, Yamamoto2010, Yamamoto2013, Meyer2001, Meyer2002} and in solutions~\cite{Muthukumar1998, Welch2001, Muthukumar2003, Zhang2007, Yamamoto1998I}, and possible precursors of crystalline lamella have been investigated~\cite{Luo2011, Meyer2002}, as well as interplay between single chain collapse and crystallization from solutions of different solvent quality~\cite{Wang2015}.
    
    Another aspect of chain folding should be mentioned here: this folding of a polymer chain in a lamella is in some sense similar to formation of a crumpled globule~\cite{Nechaev} in the coarse of a single chain collapse. The difference between a crystalline polymer lamella and a crumpled globule is the existence of a long-range ordering (both orientational and translational) of segments (stems) of a long polymer chain in a lamella and only a quite short-range ordering of folded (lamellar-like) parts in a crumpled globule. However, this analogy with crumpled globule conformation has not been raised yet in the literature except our previous work~\cite{chertovich2014crumpled}, to the best of our knowledge, although it is closely related to the intramolecular nucleation model by Hu et.al.~\cite{HuFrenkel, HuFrenkel2, Hu3}. Folded conformations of a polymer chain have been observed in simulations (coarse-grained on the united atom level), including both adjacent and random reentry~\cite{Luo2011, Yamamoto2008, anwar2013crystallization, Hu3}, although in most cases only for a single lamella and using some special techniques like self-seeding, or presence of a nucleation surface, or a template layer~\cite{Luo2011, Yamamoto2008, anwar2013crystallization, Hu3} (due to computational time limitations).
    
    In order to observe in computer simulations the appearance of folds in a crystallizing polymer chain and their distribution between different seeds of crystalline lamellae in the course of crystallization while following this process from an initial state of a homogeneous solution/melt, we need larger length and time scales, i.e., we need a coarse-graining going beyond the UA level, and this is the goal of our paper. Running ahead of the story, we have observed the nucleation and growth of lamellar seeds. 
    
   	Our purpose is to reveal a general scenario of crystallization in a particular class of polymers (long semiflexible chains) under particular conditions (poor-solvent solutions, i.e., fast precipitation accompanied by crystallization). Studying crystallization scenario means studying both the primary and secondary nucleation, both the structure and size of crystallites as well as conformational properties of single chains, i.e., we need large length and time scales. What such a CG model (beyond UA level) should take into account? It should definitely take into account the steric repulsion between CG beads (excluded volume interactions) and the intramolecular stiffness, while a particular choice of intermonomer attraction (due to Van-der-Waals interactions or specific interactions like, e.g., hydrogen bonds) is not crucial for our goals. These assumptions are based on the general understanding of crystallization mechanisms in UA models of PE and PVA~\cite{Luo2011, anwar2013crystallization, Yi2013, Yamamoto2013, Luo2014} suggesting that intermonomer attraction plays the major role only at later stages of crystallization. During initial stages of crystallization the orientational (nematic) ordering~\cite{Onsager, KhoSem} of chain segments (due to the excluded volume effect, i.e., pure steric interactions) plays the most important role, it leads to chain extension and increasing of local concentration, and finally to crystallization~\cite{anwar2013crystallization}.
    	       
    In our research we have used a very coarse-grained and popular Dissipative Particle Dynamics (DPD) technique~\cite{groot1997dissipative,espanol2017perspective} for the simulation of polymer crystallization in poor--solvent solutions and melts of semiflexible polymers. DPD allows to go to quite large time and space scales because soft potentials allow to increase the integration time step. This makes DPD considerably faster than standard MD scheme for studies of condensed polymer materials on large time scales. DPD method is very good for studying equilibrium properties, like block-copolymer microphase separation~\cite{espanol2017perspective, gavrilov2011phase, gavrilov2013phase}. Because of soft nature of used potentials there are no strong restrictions on excluded volume of beads, and this makes chains partly phantom. Therefore, the original DPD model~\cite{groot1997dissipative} is not suitable to study crystallization. However, recently it was shown~\cite{nikunen2007reptational} that it is possible to choose DPD simulation parameters to keep chains non-phantom and simultaneously use comfortably large integration step.
	In this paper, we provide DPD computer simulation of poor--solvent solutions and melts of semiflexible polymers. Chain stiffness is introduced by applying stiff spring potential on bonds between beads successive along the chain, like in the tangent hard spheres model~\cite{vega2001liquid}. This potential can reestablish steric interaction even in the model with soft-core repulsion potential, and this stimulates chain segments to undergo liquid crystalline (nematic) transition~\cite{Onsager, KhoSem} which is then the first stage of crystallization in our systems. Connectivity of monomer beads in chains, intrachain stiffness and topological restrictions (entanglements of non-phantom chains) are the three ``whales'' on whom the crystallization behavior in our model ``rests''. Although individual features of the crystallization process in a specific polymer can be different, we believe that general trends and several common regularities of melt and poor-solvent solution crystallization process should be the same in all semiflexible polymers, and that our model is able to grasp the essence of this process.
    Our model is closely related to crystallization of semiflexible polymers caused by their fast precipitation from a poor-solvent solution~\cite{Raos1997, Wang2015}, which is typical for many polymer processing schemes, including fiber formation~\cite{eichhorn2009handbook}. For such process we try to reveal some general features, both at early and later stages of crystallization, which do not depend on polymer chemical structure but are related to universal polymer properties, i.e., chain connectivity, stiffness and entanglements. 
        
    This paper is organized in a traditional way: we start with description of our model and simulation techniques, then present our results and finish with conclusions.

	\section{\label{sim-meth}SIMULATION METHODOLOGY}
	
	DPD is a method of a coarse-grained molecular dynamics with a stochastic DPD thermostat conserving total momentum and angular momentum and with soft potentials mapped onto the classical lattice Flory--Huggins theory~\cite{groot1997dissipative, espanol2017perspective}. Macromolecules are represented in terms of the bead-spring model, with particles of equal mass (chosen to be the mass unity) and equal size. One polymer bead in our model resembles a part of a polymer chain consisting of several monomer units, and one bead of a solvent includes several solvent molecules. Beads are interacting by pair-wise conservative force, dissipative force and random force:
    
	\begin{equation}
	\bm{f}_i = \sum_{i\neq j}\ (\bm{F}^b_{ij}+\bm{F}^c_{ij}+\bm{F}^d_{ij}+\bm{F}^r_{ij}) \;,
	\label{eq3}
	\end{equation}
	where $\bm{f}_i$ is the force acting on the $i$-th bead, and the summation is performed over all neighboring beads within the cut-off radius $r_c$, which is chosen to be the length unity, $r_c = 1$. First two terms in the sum are conservative forces. The term $\bm{F}^b_{ij}$ is the spring force describing chain connectivity of beads:
	
	\begin{equation}
	\bm{F}^b_{ij}=-k(r_{ij}-l_0)\frac{\bm{r}_{ij}}{r_{ij}} \;,
	\end{equation}
	where $\bm{r}_{ij} = \bm{r}_j - \bm{r}_i$, $\bm{r}_i$ is the coordinate of the $i$-th bead, $r_{ij} = |\bm{r}_{ij}|$, $k$ is the bond stiffness parameter, $l_0$ is the equilibrium bond length. If beads $i$ and $j$ are not connected by bonds, $\bm{F}^b_{ij} = 0$.
	The term $\bm{F}^c_{ij}$ is the soft core repulsion between beads $i$ and $j$:
	
	\begin{equation}
	\bm{F}^c_{ij}=\left\{
	\begin{array}{c}
	a_{ij}(r_c-r_{ij})\bm{r}_{ij}/r_{ij}, \quad \quad r_{ij}\leq r_c\\\\
	0, \quad \quad \quad \quad \quad \quad \quad \quad r_{ij}>r_c
	\end{array}
	\right.
	\;,
	\end{equation}
	where $r_c \equiv 1$, $a_{ij}$ is the maximal repulsion force between beads $i$ and $j$ for $r_{ij} = 0$. Since $F^c_{ij}$ has no singularity at zero distance, a much larger time step than in the standard molecular dynamics can be used without loosing the stability of a numerical scheme for integrating the equations of motion, and this makes it possible to access larger time scales when complex polymeric systems are studied~\cite{groot1997dissipative,espanol2017perspective}. Other constituents of $\bm{f}_i$ are random force $\bm{F}^r_{ij}$ and dissipative force $\bm{F}^d_{ij}$ acting as a heat source and surrounding media friction, respectively~\cite{groot1997dissipative}. The parameters for these forces are: noise parameter $\sigma = 3$, friction force parameter $\gamma=4.5$. More detailed description of our simulation methodology can be found elsewhere~\cite{gavrilov2011phase,gavrilov2013phase}. 
    
	We study systems with different polymer volume fractions $\varphi = 20\%$, $50\%$, $70\%$, $80\%$, $90\%$, $95\%$ and $100\%$. This total polymer volume fraction is constant during the simulation, but the local polymer concentration can change significantly. Since the DPD scheme uses explicit solvent particles, for systems with $\varphi < 1$ the rest of simulation box is filled by solvent beads. The total number density of DPD particles in our systems was $\rho=3$. The repulsion parameter between a polymer particle and a solvent particle, $a_{ps}$, was chosen to be larger than polymer--polymer interaction parameter, $a_{pp}$, while their difference was chosen to be  $\Delta a = a_{ps}-a_{pp}=10$, unless otherwise specified in the text. Such condition corresponds to a poor solvent case, and the Flory-Huggins parameter of polymer-solvent interaction can be calculated as $\chi_{ps}=0.306\Delta a$ (see Ref.~\cite{groot1997dissipative}) and occurs to be $\chi_{ps} \approx 3$ (in our simulations $k_BT=1$). This mimics the situation of a fast polymer precipitation from solution with simultaneous crystallization, which is typical for many polymer processing schemes, including fiber formation~\cite{eichhorn2009handbook}.

	The use of soft volume and bond potentials leads to the fact that the chains are formally ``phantom'', i.e., their self-crossing can happen in three dimensions. The phantom nature of chains can affect both the equilibrium properties (e.g., the phase behavior of the system or relationships between the average radius of gyration of a coil and the number of units in the coil) as well as dynamic properties (while the dynamic/kinetic properties can be affected more strongly). However, it greatly speeds up the equilibration of the system. As regards the dynamic properties, it was shown that the original DPD method is consistent with the Rouse dynamics~\cite{spenley2000scaling,lahmar2007influence} which is relevant only for ideal polymers and non-entangled polymer melts. However, if studying crystallization behavior, that requires consideration of explicit entanglements between chains due to steric interactions, it is necessary to introduce some additional forces that forbid the self-intersection of the chains. These forces are usually quite cumbersome and considerably slow the computation. There are several methods developed to avoid (or at least to reduce significantly) the bond crossings in CG models (see, e.q., the review~\cite{Gooneie-Polymers2017}). Nikunen et al.~\cite{nikunen2007reptational} described a method for keeping chains to be non-phantom in DPD simulations without any additional forces. We have used this quite simple and fast method~\cite{nikunen2007reptational} which has been proven to reproduce entanglements and polymer reptational dynamics reasonably well~\cite{espanol2017perspective, Karat-Winey-2013, chertovich2014crumpled, nikunen2007reptational}.
	
	\begin{figure}[htbp]
	\includegraphics[width=0.4\linewidth,keepaspectratio]{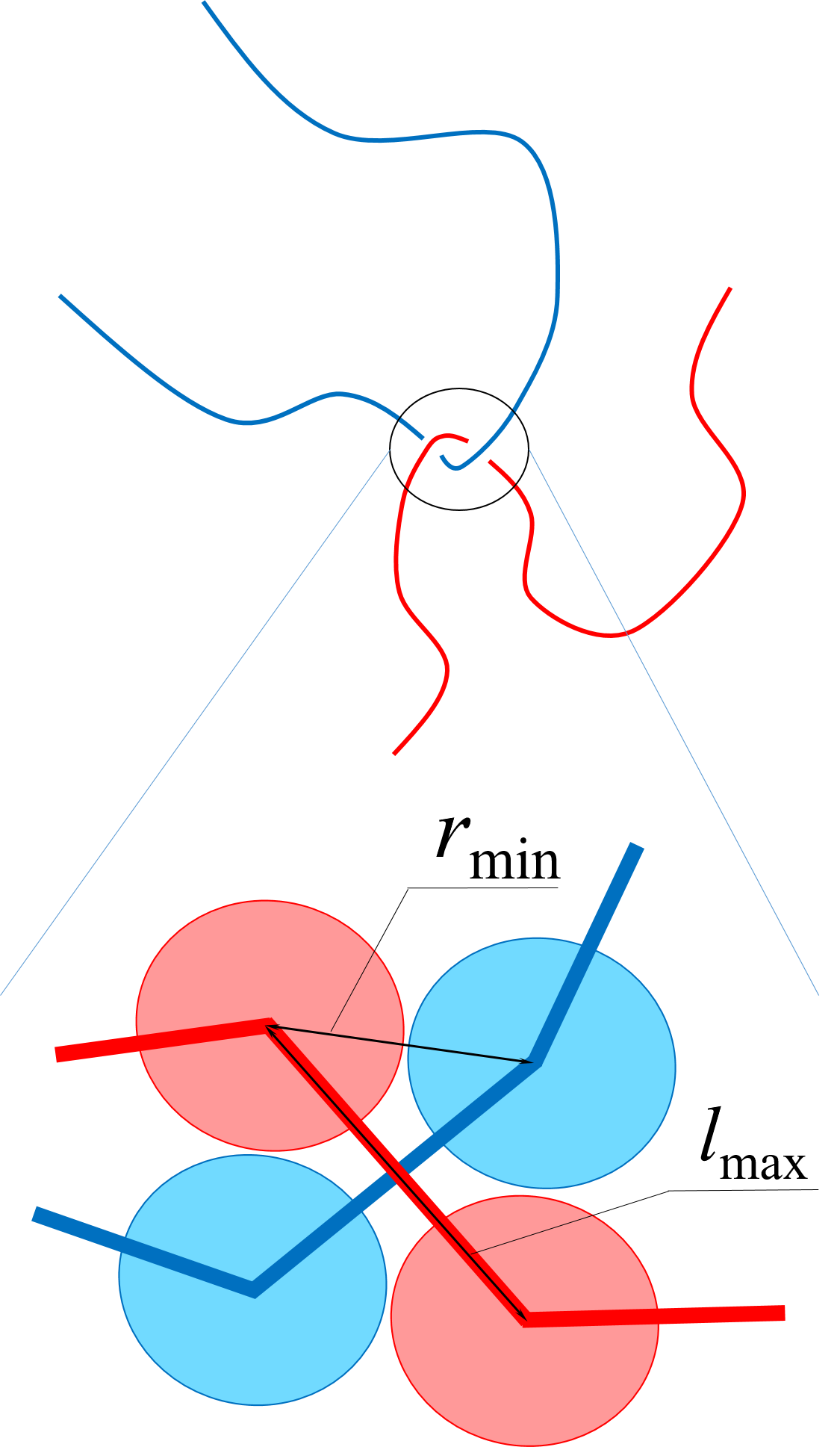}
	\caption{Schematic representation of two chains to check bonds crossing conditions.}
	\label{like_nik}
	\end{figure}	
	
	This method is based on geometrical considerations (see Fig. \ref{like_nik}). If the distance between any two beads which are not neighbors along the chain cannot be smaller than $r_{min}$, every bead in the system effectively has an excluded volume with radius $r_{min}/2$. If each bond along the chain can have the length not larger than the maximum length $l_{max}$, the condition of non-phantom chains is satisfied if inequality $\sqrt{2}r_{min} > l_{max}$ holds in the course of simulations.~\cite{nikunen2007reptational} Although particles in DPD are formally point-like, they still have some excluded volume due to the presence of the repulsive potential governed by the $a_{ij}$ value. Similarly, due to the bond potential all bonds can have some maximal possible length~\cite{nikunen2007reptational}. In our study, we have chosen $a_{pp} = a_{ss} = 150$, $a_{ps} = 160$, $l_0 = 0.2$, $k = 150$, and with this set of parameters the probability of chain intersection events in our simulations is negligibly small, see Fig. S1 in Supplementary Material. Recently, it was also shown that if the chain crossing events are rare, the entanglement behavior in polymer melt is recovered.~\cite{chang2019can} In addition, the small value of equilibrium bond length $l_0 = 0.2$ forces beads, which are the neighbors along the chain, to be located considerably closer to each other than non-linked beads. Repulsive volume interactions between beads $i$ and $i \pm 2, i \pm 3...$ provide effective chain stiffness, similar to that in the tangent hard spheres model~\cite{vega2001liquid}. This leads to some ``hardening'' of soft interaction potential (due to quite large values of $a_{ij}$-parameters and bond stiffness $k$ as well as small value of $l_0$), but we were still able to use a quite large integration time step $\Delta t=0.02$. Therefore, we  have a solution of self-avoiding semiflexible chains, and such a system has a tendency to undergo lyotropic nematic ordering transition~\cite{Onsager, KhoSem} at high polymer concentrations due to steric interactions only, and this orientational ordering is often the first stage of a possible crystallization transition~\cite{anwar2013crystallization, anwar2015crystallization, MarkinaPI}. The obtained average distance between linked monomer units is $\langle l \rangle = 0.48$ (in units of $r_c$), the chain persistence length $b$ is about $2.3$ monomer units, i.e., $\tilde{b} = b \cdot \langle l \rangle = 1.1$ in units of $r_c$, as we have estimated using the bond autocorrelation analysis~\cite{gowers2016mdanalysis,michaud2011mdanalysis} (for calculation of persistence length a single chain was simulated in a $\theta$-solvent with $\chi=0.5$ and the data has been averaged over 5 independent runs). Note, that the intramolecular stiffness is actually not large in our model, however, the additional stiffening of chains (increasing of the persistence length) takes place due to nematic ordering~\cite{Ivanov2014}.
	
	We have used rather long chains of the length $N=10^3$ monomer units. The initial conformations of the system were prepared as sets of Gaussian chains (random walks) with bond length $l = 0.48$ and with the desired polymer volume fraction $\varphi$. Then, the remaining volume was randomly filled with the solvent particles until the total number density $\rho=3$ was reached. Each system was equilibrated for $10^6$ DPD steps in $\theta$-solvent. At this stage of the simulation, the systems were pre-structured. In the case of a melt, crystallization simply was started. Thereafter, the quality of solvent was instantly changed to poor-solvent conditions and the rest of each simulation was $10^8$ DPD steps.
	
	The simulation box was set to be cubic with the side size $L_{box}=50 r_c$, with periodic boundary conditions in all directions. There were totally 375 000 DPD beads in simulation box, and the maximum number of polymer chains was equal to $N_{chains}^{(max)}=375$ in the case $\varphi=1$. The average end-to-end distance of a Gaussian chain with $N = 10^3$ and $l = 0.48$ is equal to $R =l\sqrt{N}\approx15$, and for semiflexible chain it becomes larger $R=2bl\sqrt{N/2b}\approx32.5$, which is still smaller than the box side (instead of $l$ we have used the Kuhn segment which is twice the persistence length $b$). Thus, on average each polymer chain does not interact with itself via periodic boundaries (or at least possible effects from such interactions are negligibly small; this is also true even when a percolating cluster appears in our simulation box because a percolating cluster consists of several chains, not of a single chain). Despite all physical parameters which we have monitored varied smoothly with time and polymer volume fraction $\varphi$, we have performed simulations of $5$ independent system for each $\varphi$-value (note also, that the parameters related to single chain properties are self-averaging, i.e., they can be averaged over all chains in a simulation box). For calculations we have used our own original domain-decomposition parallelized DPD code~\cite{gavrilov2011phase, gavrilov2013phase}. The maximum simulation time was $10^8$ DPD steps for each polymer volume fraction $\varphi$.

	\begin{figure}[htbp]
		\includegraphics[width=0.7\linewidth,keepaspectratio]{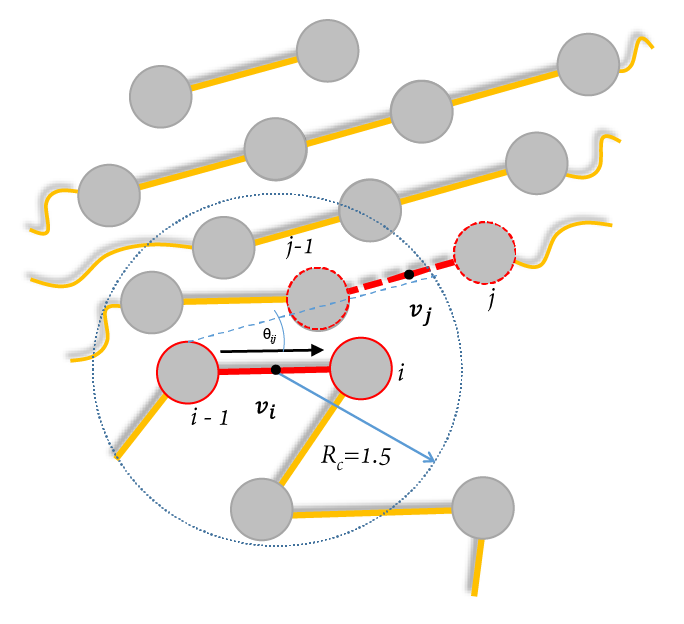}
		\caption{Schematic illustration of the crystallinity criterion for a polymer bond $\bm v_i$. }
		\label{schema}
	\end{figure}

	To characterize the system morphologies, we implement the following two-stage cluster analysis. On the first stage, we consider bonds $\bm v_i$ connecting two successive beads (monomer units) $i-1$ and $i$ along the chain (Fig. \ref{schema}), while coordinates of centers of bonds are chosen to be the coordinates of bonds, and directions of bonds are vectors between two successive beads. We determine which bonds belong to the crystalline fraction, and then we perform cluster analysis to find crystallites composed from these ``crystalline'' bonds. To recognize whether a particular bond $\bm v_i$ belongs to the crystalline phase we use the following rules (see Fig. \ref{schema}).

	\begin{enumerate}
		\item For each bond $\bm v_i$, we determine all neighboring bonds $\bm v_j$ as bonds inside a sphere with the center at the center of bond $\bm v_i$ and radius $R_c = 1.5r_c$. In the system with polymer volume fraction $\varphi=1$, each bond has $\sim 18$ neighbors.
		\item Then, we define the number of neighboring bonds $\bm v_j$, which are collinear with the selected one, $\bm v_i$. For this goal, we calculate the angle $\theta_{ij}$ between bonds $\bm v_i$ and $\bm v_j$ and use the following collinearity threshold criteria: two bonds $\bm v_i$ and $\bm v_j$ are collinear if the angle between them is less than 17 degrees, $\theta_{ij} < 17$ degrees. 
		\item Finally, we calculate the ratio of the number of collinear neighbors to the total number of neighbors. Bond $\bm v_i$ is marked ``crystalline'' if this ratio is greater than $0.4$, otherwise it is marked ``amorphous'', and this is our bond crystallinity criteria (this procedure is quite similar to the crystallinity criteria used previously in MD simulations of crystallization in alkanes~\cite{anwar2013crystallization}).
	\end{enumerate}
   After such labeling, we perform the standard clustering analysis~\cite{allen1989computer} using cut-off radius $R_c = 1.5r_c$ to find the clusters formed by neighboring ``crystalline'' bonds.

	At the second stage of our analysis of crystalline clusters, we turn to the consideration of the beads and mark a bead ``crystalline'' if it is either the beginning or the end bead of at least one crystalline bond, otherwise it is marked ``amorphous''. This procedure transforms the system of {\it bonds} into the system of {\it beads}. However, after the first stage several ``amorphous'' beads were still located inside some crystalline clusters (although they did not belong to those crystalline clusters). To decide, whether an ``amorphous'' bead in the neighborhood to some crystallite, i.e., being located within the cutoff radius $R_c = 1.5r_c$ from any bead of this crystallite, still can be included in this crystallite, we define the direction of a bead $i$ as the vector from the bead $(i-1)$ to the bead $(i+1)$ and  check the following criteria:
	\begin{enumerate}
		\item Calculate the director $\bm{D_c}$ for each cluster, i.e., the unit vector corresponding to the preferred orientation of the beads in a crystallite (the normalized sum of all beads in a cluster).
		\item Compare the directions of the cluster director $\bm{D_c}$ and a neighboring ``amorphous'' bead. If the angle $\alpha$ between the cluster direction and the bead direction is less than $\alpha < 20$ degrees, the bead is marked ``crystalline'' and added to this cluster, otherwise the bead stays ``amorphous''. 
	\end{enumerate}

	This additional second stage of analysis of beads initially marked as ``amorphous'', is aimed to avoid splitting of rod-like segments (stems) inside crystalline clusters (which we also call crystallites or lamellae), that can happen because of coordinates/directions fluctuations due to the soft nature of excluded volume potential in DPD. 

	In our simulations, we have monitored the average size and the number of crystallites, as well as the degree of crystallinity as function of time and polymer volume fraction. To characterize polymer chain conformations, we have calculated the dependence of the average squared spatial distance between two monomer units on the number of monomer units between them along the chain, $R^2(n)$, separately for the segments belonging to crystalline and amorphous fractions, respectively. Such dependence has been proved to be very useful to analyze polymer conformational state at different length scales~\cite{lifshitz1978some,chertovich2014crumpled}. We have also studied the length distribution of crystalline segments or stems, i.e., the strongly extended parts of polymer chains inside crystallites, for different polymer concentrations.

	\section{\label{res-n-disc}RESULTS AND DISCUSSION}
    
	\begin{figure}[htbp]
	\centering
		a) \includegraphics[width=0.45\linewidth]{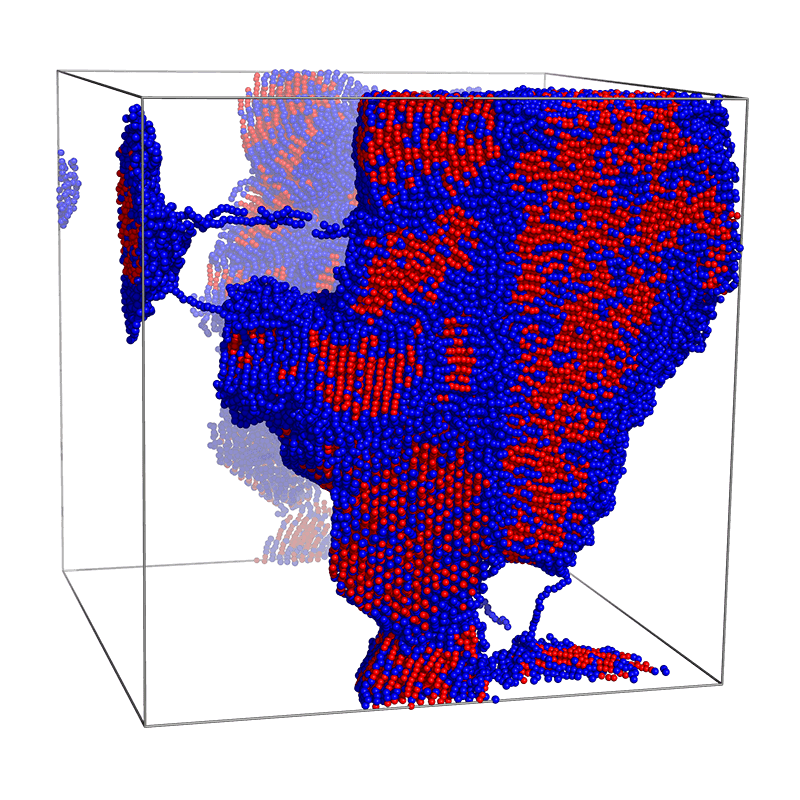}
		b) \includegraphics[width=0.45\linewidth]{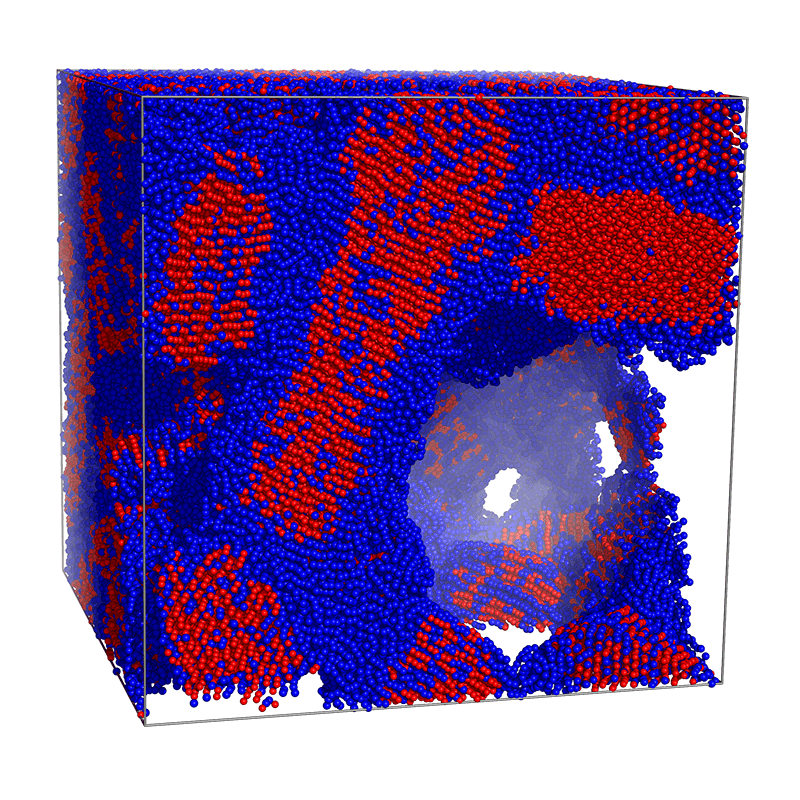} \\
		c) \includegraphics[width=0.45\linewidth]{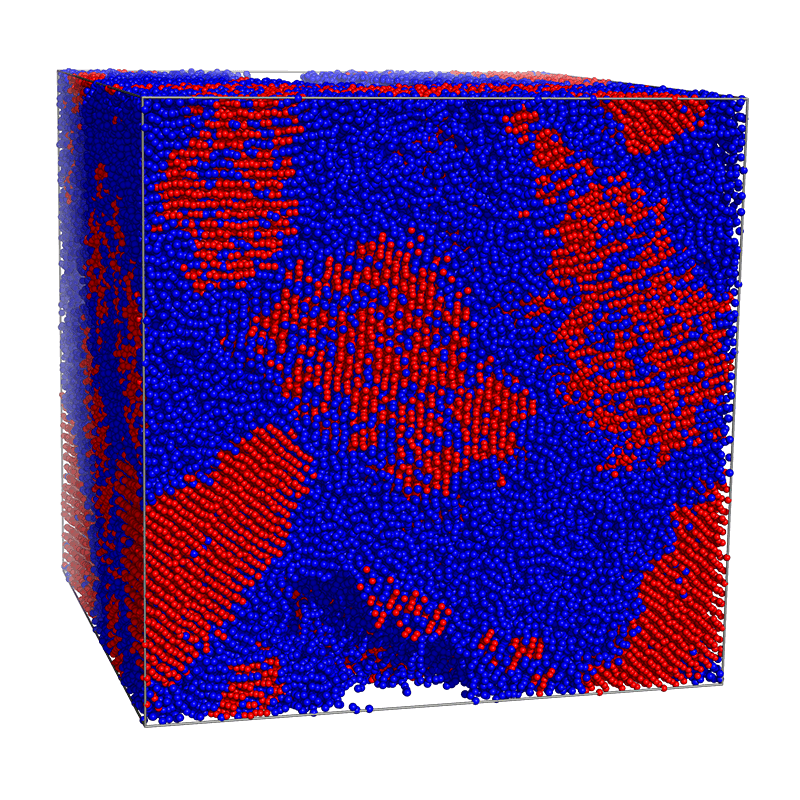}
		d) \includegraphics[width=0.45\linewidth]{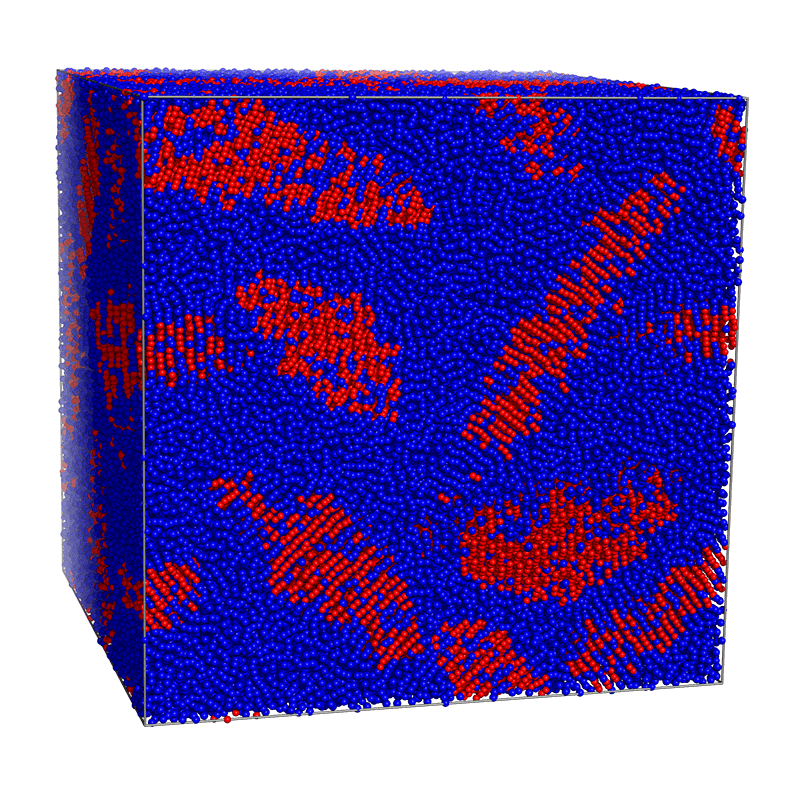}
	\caption{Snapshots of several systems with different polymer volume fraction $\varphi$ at the end of equilibration, $t = 10^8$ DPD steps: $\varphi=0.2$ (a),  $\varphi=0.7$ (b), $\varphi=0.9$ (c), $\varphi=1$ (d). Only polymer beads are shown (``crystalline'' beads are red, ``amorphous'' beads are blue), while solvent beads are hidden.}
	\label{confs}
	\end{figure}

	Final morphologies after long equilibration are shown in Fig.~\ref{confs} for several systems having different polymer volume fraction $\varphi$. In Fig.~\ref{confs}, solvent particles are hidden, and red and blue colors represent ``crystalline'' and ``amorphous'' beads, respectively. It is easy to recognise crystallites (red) separated by solvent (transparent) or amorphous polymer (blue) regions. The linear size of a typical crystallite is smaller than the simulation box size, so that a single crystallite does not interact with itself through periodic boundaries (a possible case of percolating crystallites is discussed below). Visually, we can get an impression that the crystallite size seems to have the largest value for polymer volume fraction $\varphi$ somewhere between 70 \% and 95 \%.

\begin{figure}[htbp]
		\centering
			a) \includegraphics[width=0.45\linewidth,keepaspectratio]{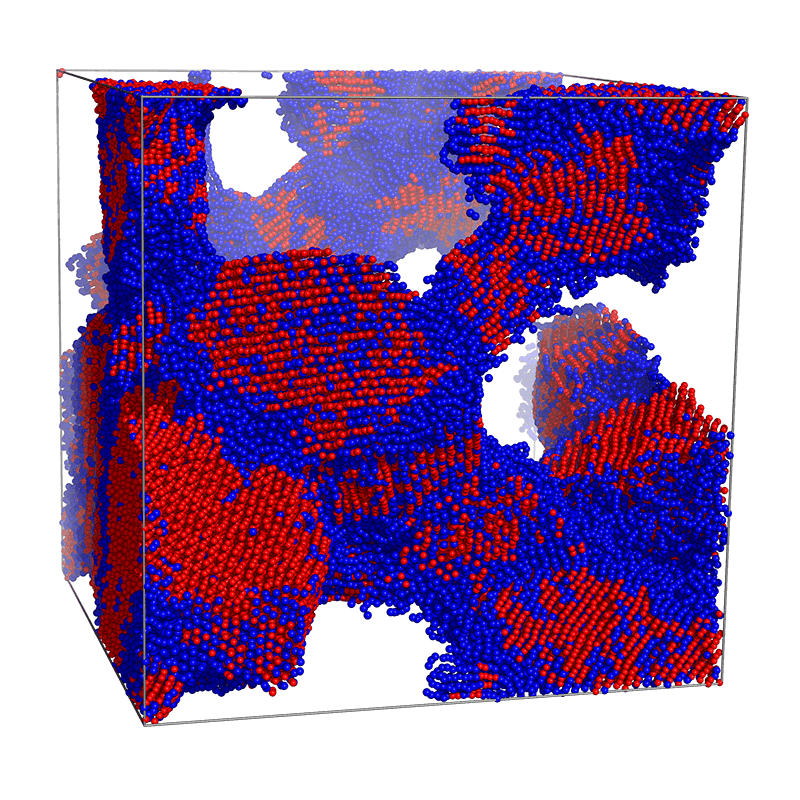}
			b) \includegraphics[width=0.45\linewidth,keepaspectratio]{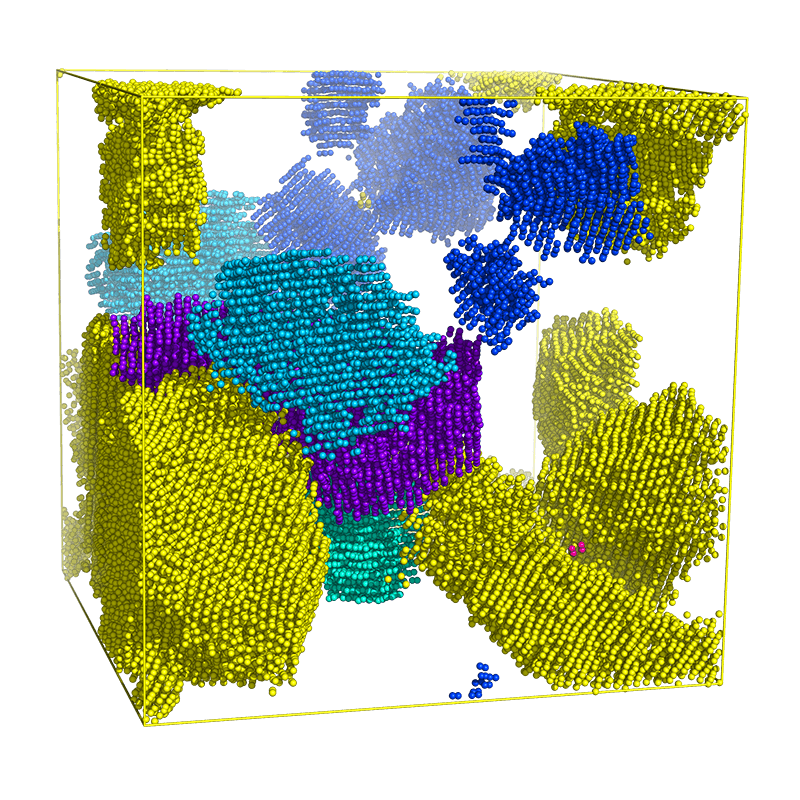} \\
			c) \includegraphics[width=0.45\linewidth,keepaspectratio]{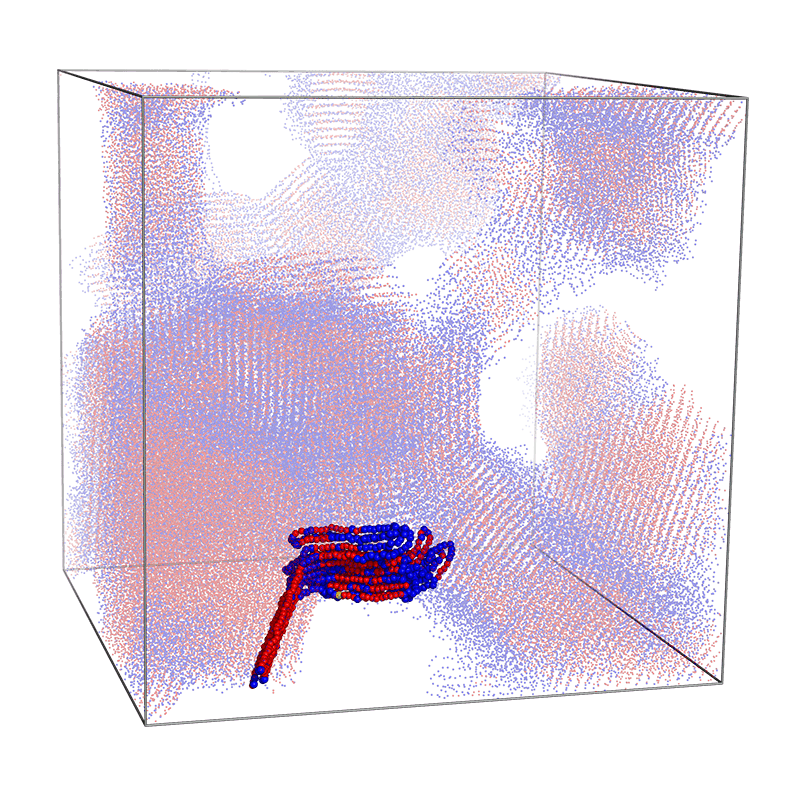}
			d) \includegraphics[width=0.45\linewidth,keepaspectratio]{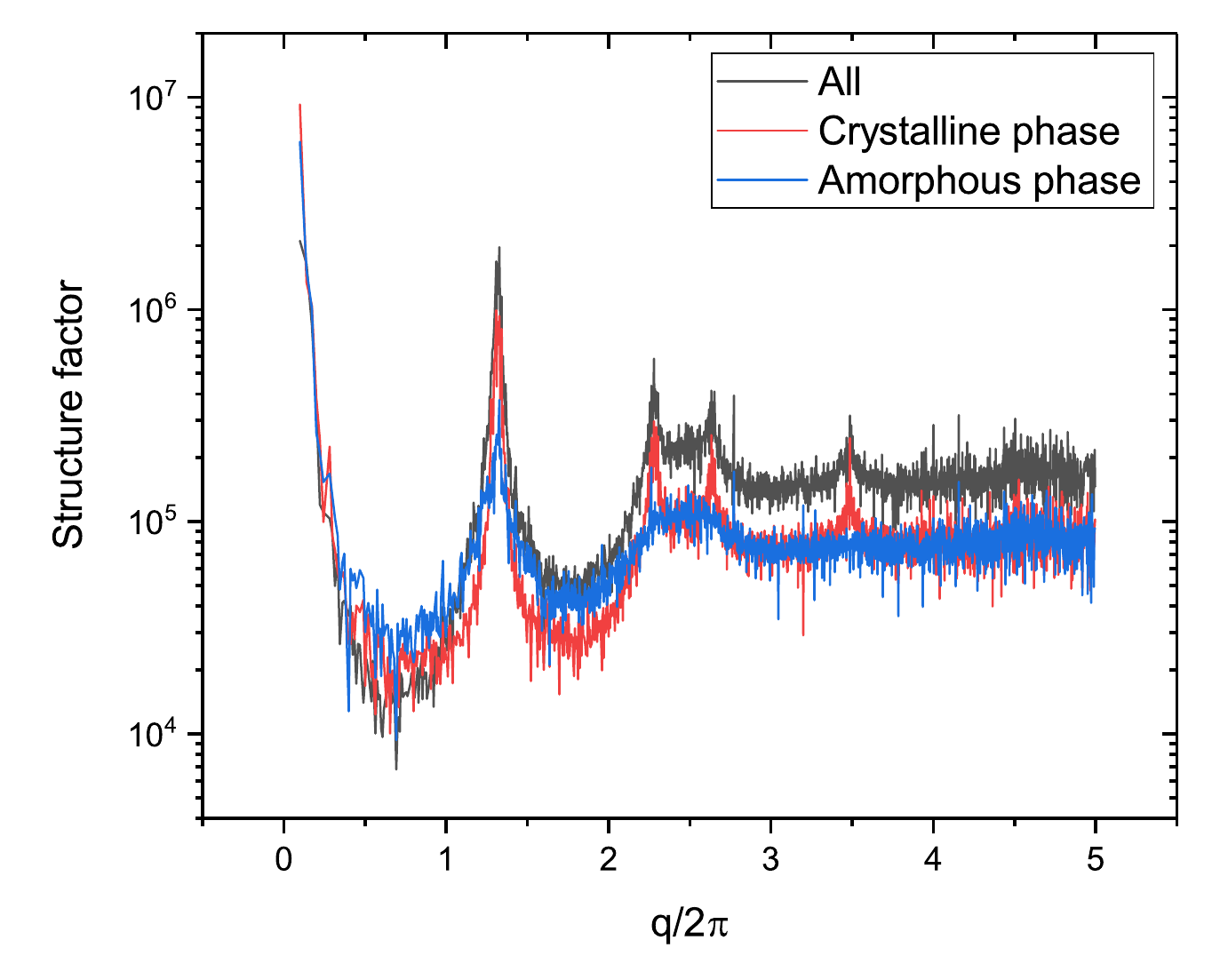}
		\caption{
        \textbf{(a)} Snapshot of a system with 50 \% polymer volume fraction (some intermediate morphology in the course of equilibration): red beads are ``crystalline'' and blue beads are ``amorphous'', solvent beads are not shown.
		\textbf{(b)} The same system after both stages of the cluster analysis: different colors correspond to different crystalline clusters, and ``amorphous'' beads are not shown.
		\textbf{(c)} The conformation of a randomly chosen single polymer chain from (a).
		\textbf{(d)} The static structure factor for the polymer beads (black) and  separately crystalline (red) and amorphous (blue) phases from (a).}
		\label{cryst_clast_single_sq}
	\end{figure}

	First $10^6$ DPD steps represent the system behavior in $\theta$-solvent, then the quality of solvent was instantly changed to poor with Flory-Huggins parameter $\chi=3$. Fig.~\ref{cryst_clast_single_sq} shows a more detailed view of a typical system with 50\% polymer volume fraction at some intermediate time of $10^7$ DPD steps, including its crystallization behavior and cluster structure. In Figure \ref{cryst_clast_single_sq}a one can see the polymer--solvent separation as well as the separation of the polymer--rich phase into crystalline and amorphous sub-phases. The crystalline sub-phase is comprised of several crystallites (see Figure \ref{cryst_clast_single_sq}b), which could be easily found by our two-stage clustering analysis described above. These crystallites can have quite complex surfaces and very different sizes. Analysis of maxima in static structure factor for polymer (see Figure \ref{cryst_clast_single_sq}d) confirms that chain parts (stems) in crystallites have hexagonal packing~\cite{Meyer2002}. This corresponds to the rotator phase. A single chain may contribute to several crystallites (e.g., the chain shown in Figure \ref{cryst_clast_single_sq}c contributes to two different crystallites). In general, the crystallite structure is similar to that obtained, e.g., for another model by means of MD simulation~\cite{Meyer2002}. Thus, a preliminary conclusion here is that our model and the DPD simulation scheme keep the chains to be non-phantom (which is crucial for polymer crystallization), reproduce reasonably good some general features and are suitable for studying crystallization in polymers on quite large time and length scales (due to soft potentials and large integration step). 

\begin{figure}[htbp]
	a) \includegraphics[width=0.95\linewidth]{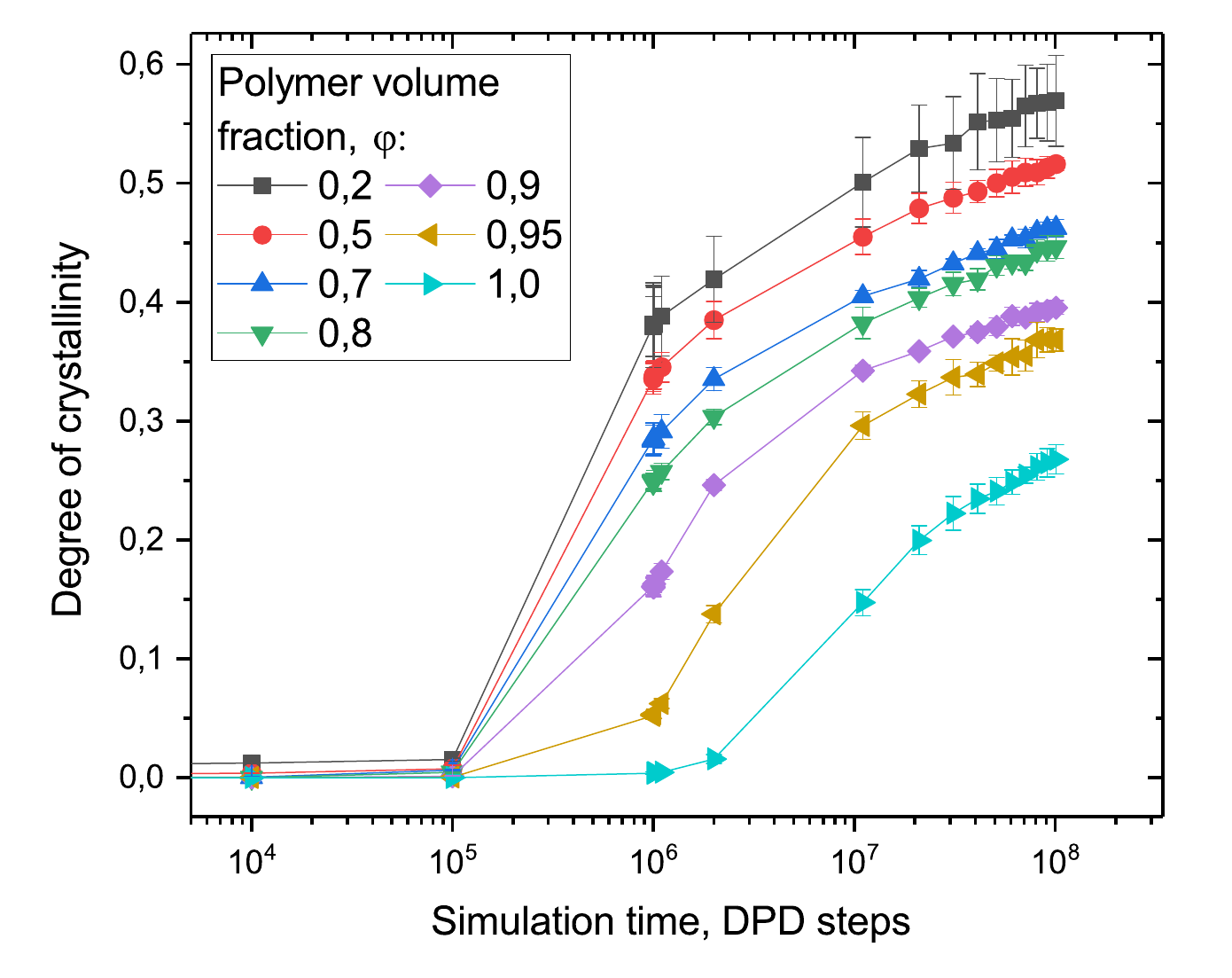} \\
	b) \includegraphics[width=0.95\linewidth]{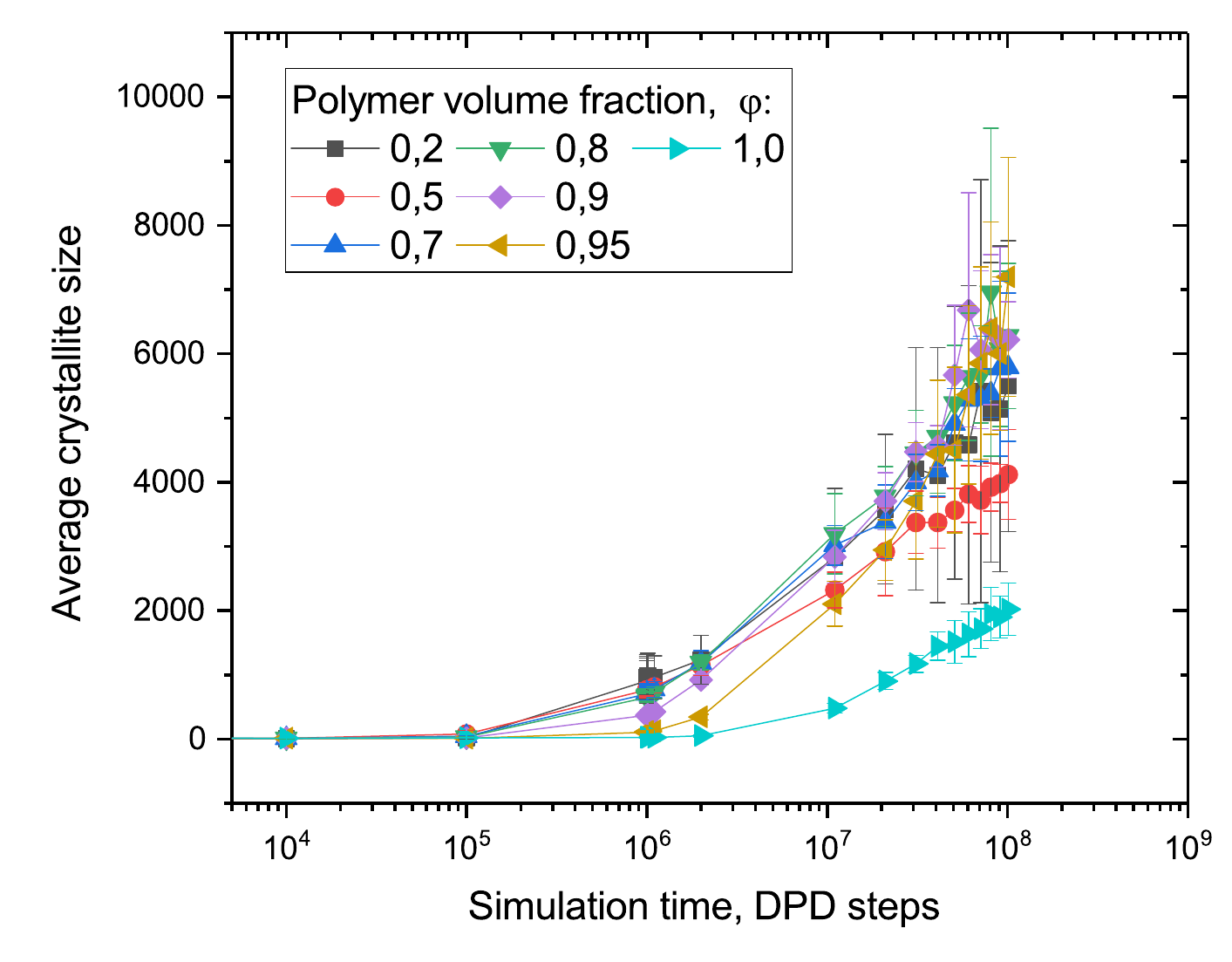} \\
	c) \includegraphics[width=0.95\linewidth]{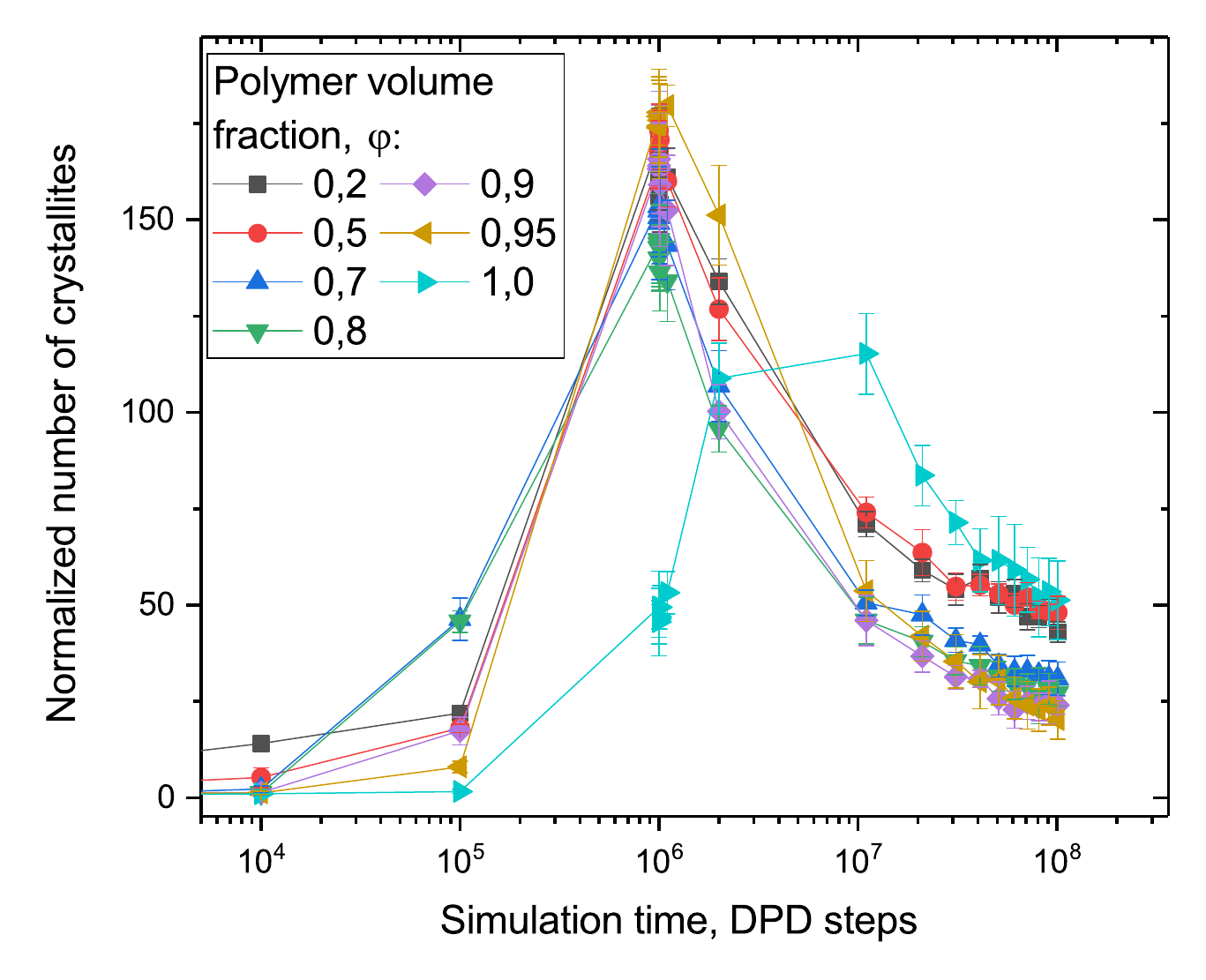}
	\caption{Time dependencies of the degree of crystallinity (a), of the average crystallite size (b), and of the normalized number of crystallites (c) for systems with different polymer volume fraction (shown in the legend).}
	\label{f_cryst_fraction}
\end{figure}
	
\begin{figure}[htbp]
\centering
	a) \includegraphics[width=0.95\linewidth]{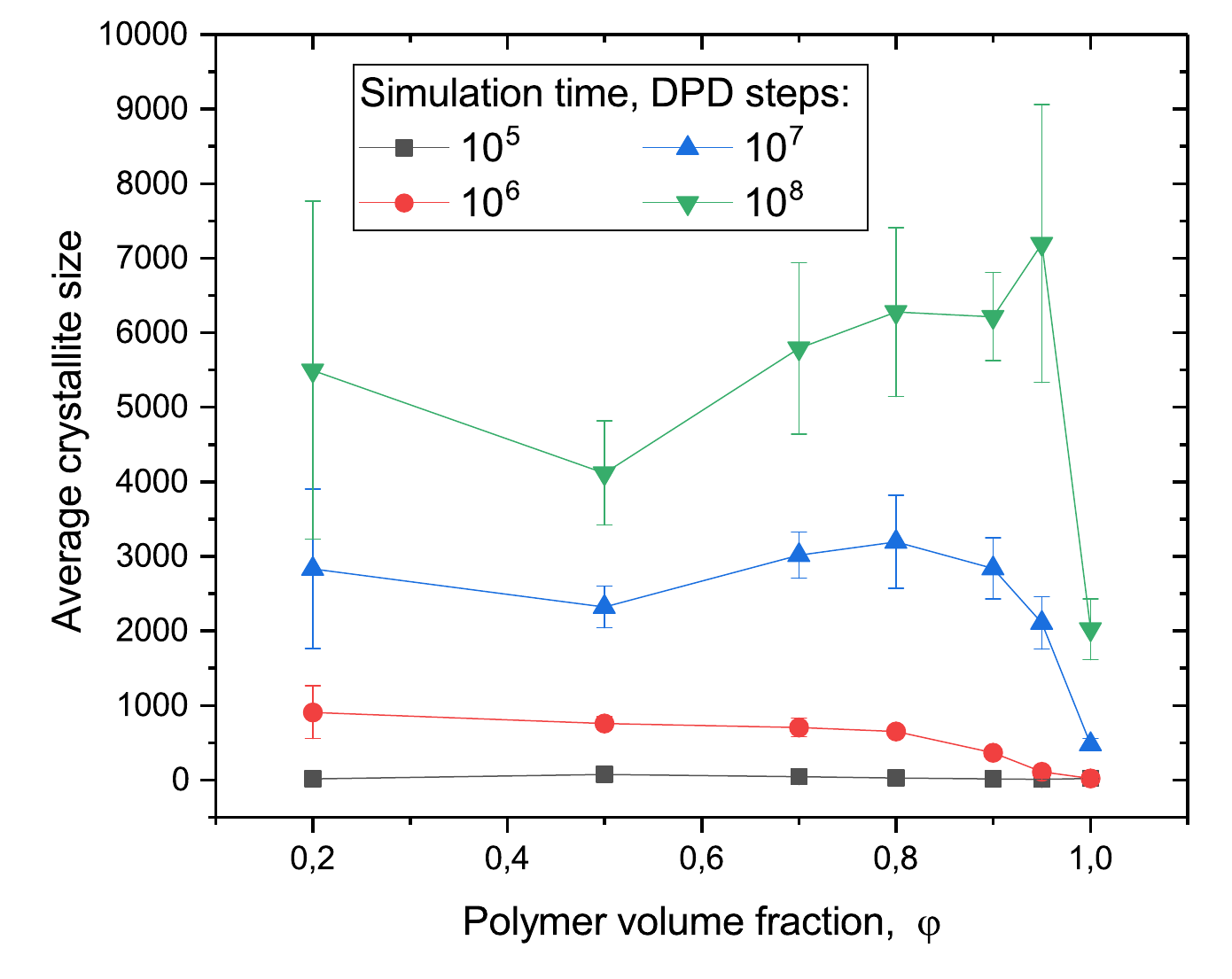} \\
	b) \includegraphics[width=0.95\linewidth,keepaspectratio]{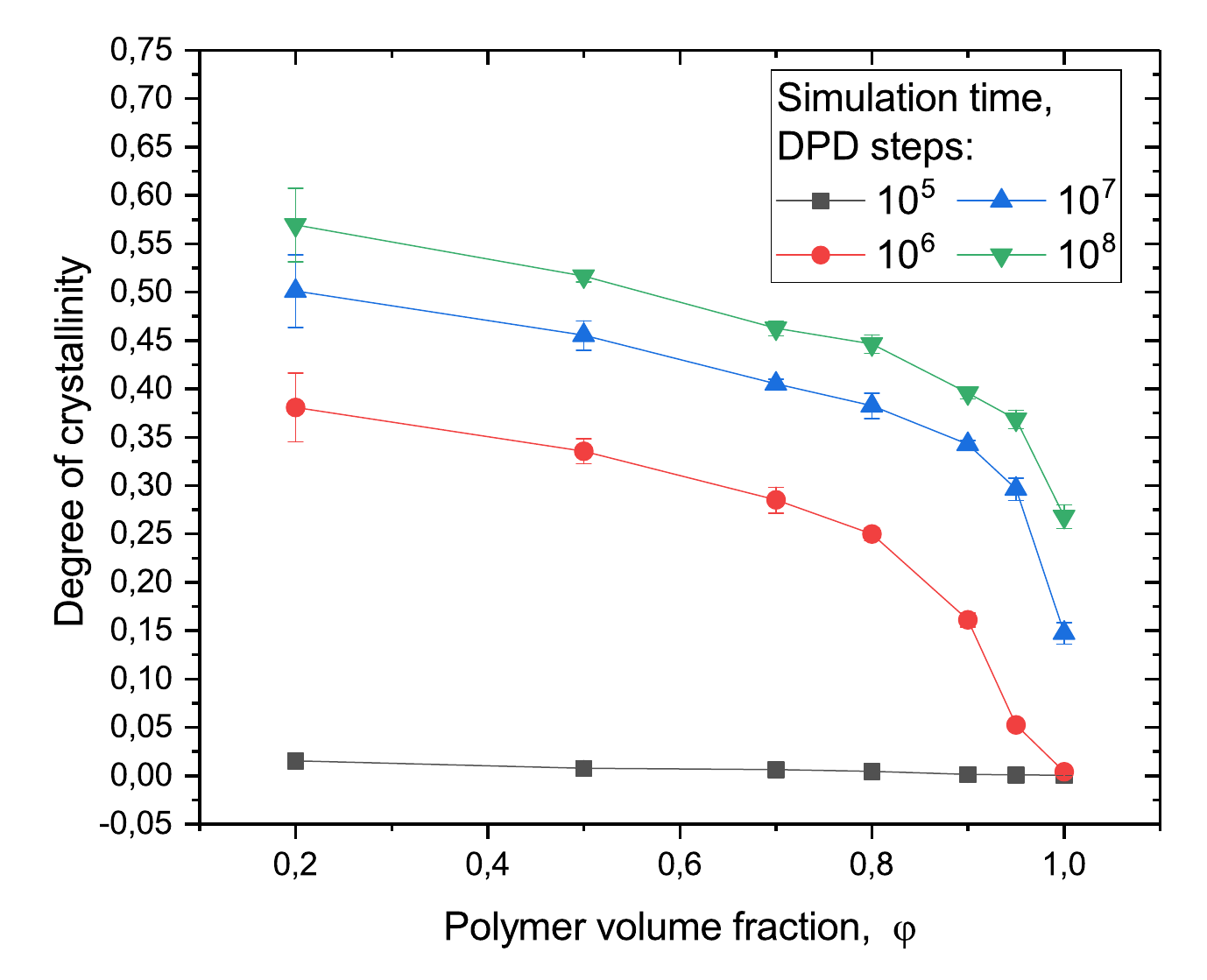}
	\caption{Dependence of the average crystallite size (a) and the degree of crystallinity (b) on polymer volume fraction at different times (in DPD steps, shown in the legend) during the crystallization process. }
	\label{fig:acs-n-cf}
\end{figure}

	Figs. \ref{f_cryst_fraction} and \ref{fig:acs-n-cf} present physical quantities which are main characteristics of the crystalline sub-phase -- the average size of crystallites (in beads), the normalized number of crystallites (i.e., the number of crystallites divided by the polymer volume fraction $\varphi$), and the degree of crystallinity (i.e., the number of crystalline beads divided by the total number of polymer beads) -- at different times and for different polymer volume fractions. 
	We indicate error bars in these figures to reveal the significance of the results. We use lines connecting the data points without any smoothing, i.e., just as a guide to the eye.
	
	In Fig. \ref{f_cryst_fraction}a one can observe the process of crystallization in logarithmic timescale -- the degree of crystallinity starts to increase quite rapidly after about $10^5$ DPD time steps for small polymer volume fractions $\varphi$ and after about $10^6$ DPD time steps for large polymer volume fractions $\varphi$. The time when this initial growth of the degree of crystallinity has a maximal slope coincides with our estimation for the local crystallization time $\tau$, i.e., the time of formation of maximal number of crystallites, from Fig. \ref{f_cryst_fraction}c. This time $\tau$ depends on $\varphi$ which is due to preliminary structuring and different speed of the polymer--solvent separation. After this initial relatively rapid increase of the degree of crystallinity, a steady-state regime of logarithmically slow crystallization process starts at times about $10^7$ DPD time steps. The kinetic reason for this slowing-down is the network of entanglements in the concentrated polymer solution (or even melt after separation from the solvent), so that the full relaxation time for such system is expected to be comparable with reptation time~\cite{DoiEdwards}, i.e., being of the order $\tau N^{3}$. Because in our simulations the polymer length is $N=10^3$, there is no hope to reach a true thermodynamic equilibrium in such systems, both in simulations and in real experiments (crystallizable polymers always stay semi-crystalline), and we can only discuss properties of some quasi-equilibrium steady-state system with slow evolution toward morphologies with higher degree of crystallinity. For our model, we observe a quite interesting feature: the crystallization speed is almost the same with good accuracy in the steady-state regime for systems with $\varphi$ from 20\% to 95\%. This is probably because the entanglements length $N_e$ is similar for those polymer systems (see also Fig.S2 in Supplementary Material for time evolution of $N_e$). These systems were calculated without the stage of preliminary structuring in $\theta$-solvent for $10^6$ DPD steps. The calculation of $N_e$ was carried out in a similar manner to the procedure developed by M. Kr\"oger's group~\cite{kroger2005shortest,shanbhag2007primitive,hoy2009topological,karayiannis2009combined} (this method gives an average value of $N_e$ for all chains in a system). We note here that in our simulations we did not observe considerable increase of $N_e$ during the whole simulation time (except the very short time in the beginning). This observation is consistent with the argument for a long reptation time given above. Entanglement length increases (i.e., the chains become less entangled) quite fast only on the times smaller than 100 DPD steps (i.e., only in the beginning of simulations after we abruptly make the solvent very poor), and afterwards it stays almost constant for all systems. The ``time of disentanglement'' is 4-5 orders of magnitude smaller than the local crystallization time $\tau$. In the initial configurations, the entanglement length is $N_e \approx 10$ for $\varphi=1$ and $0.9$ and $N_e \approx 70$ for $\varphi=0.5$. During the first 100 DPD steps it increases to $N_e \approx 30$ for $\varphi=1$, $N_e \approx 40$ for $\varphi=0.9$ and $N_e \approx 200$ for $\varphi=0.5$. At the end of simulation ($10^8$ DPD steps), $N_e$ becomes slightly larger in concentrated systems and almost does not change in dilute systems: $N_e \approx 40$ for $\varphi=1$, $N_e \approx 60$ for $\varphi=0.9$ and $N_e \approx 200$ for $\varphi=0.5$.
	
    Time evolution of the average crystallite size (measured in beads) shows a rather fast growth on times less than $10^6$ DPD steps and much more slow growth on larger times (Fig. \ref{f_cryst_fraction}b). Interestingly, the largest average cluster size is achieved for polymer volume fraction in the range $\varphi \in [0.7, 0.95]$ (see more discussion below). The data for average cluster size (used in Fig. \ref{f_cryst_fraction}b and in Fig. \ref{fig:acs-n-cf}a below) were obtained using the small size cutoff value equal to 5 in the averaging of the cluster size distribution.
        
	Figure \ref{f_cryst_fraction}c shows the total number of determined crystallites divided by the polymer volume fraction $\varphi$ and gives more insight about the crystallization process. On small times, we observe a sharp increase of the number of crystallites. One can see a well-defined maximum (local crystallization time $\tau$) around $10^6$ and $10^7$ DPD steps for systems with small and large polymer volume fractions, respectively. A pronounced decrease of the number of crystallites is consistent with the fast growth of average crystallite size. The initial increase represents the nucleation process, when many small crystallites (crystalline seeds) are formed in the system, and this stage of crystallization starts to saturate at the time moment when the overall degree of crystallinity starts to increase, see  Fig. \ref{f_cryst_fraction}a. After that moment the number of crystallites starts to decrease both due to the process of merging of small crystallites into larger ones and due to adding new chains stems from amorphous sub-phase to crystallites. Merging of small crystallites occurs most probably due to filling the space between two crystallites by new chain stems from amorphous phase which becomes crystalline but not due to diffusion of crystallites towards each other. To summarize, at the late stages of crystallization process (after $10^7$ DPD steps) the degree of crystallinity and the average size of crystallites are increasing while the number of crystallites is decreasing. Increasing of average crystallite size is more slow on the time interval $10^7 - 10^8$ DPD steps where one can even suspect logarithmic dependence. It seems that the length and time scales available in our simulations allow observation of many crystallite seeds in the simulation box and their growing and merging. We would like to mention that in the movie (see Fig. S3 and movie in Supplementary Material) of the system evolution we have observed both the addition of chain stems to the lateral faces of growing crystallites and the diffusion of chain parts inside crystallites trough their end faces, as it was also observed previously in MD simulations of UA models~\cite{anwar2013crystallization, anwar2015crystallization, Luo2011, Yamamoto2013}. In our opinion, our observations are in some content also similar to a two stage crystallization scheme~\cite{Strobl2000}: first, some seeds (precursors) of crystallites occur in a polymer melt, and secondly, these precursors rapidly aggregate to form lamellae. Interestingly, that during the second regime all systems have very similar behavior, and we can suspect even the existence of a universal power-law decrease of the number of crystallites at late stages of crystallization. 
    
    In Fig. \ref{fig:acs-n-cf}a we observe another interesting feature: the dependence of the average crystallite size (in beads) on polymer volume fraction $\varphi$ increases at intermediate values $\varphi \in [0.7, 0.95]$. In Fig. \ref{fig:acs-n-cf}b, the degree of crystallinity for different $\varphi$ is also permanently growing in the dependency of time but it is monotonically decreasing in the dependency of $\varphi$ at any time point. We suppose that this feature originates from the competition of two opposite processes that limit the crystallite growth: (i) at small concentrations there is a lack of surrounding polymer material, while (ii) at high concentrations the entanglement effect from surrounding chains is so strong that it prevents the segments diffusion on large distances as well as the local orientational rearrangement of polymer segments in amorphous regions (these are two processes, which are necessary for merging of small crystallites into larger ones).

\begin{figure}[htbp]
		a) \includegraphics[width=0.95\linewidth]{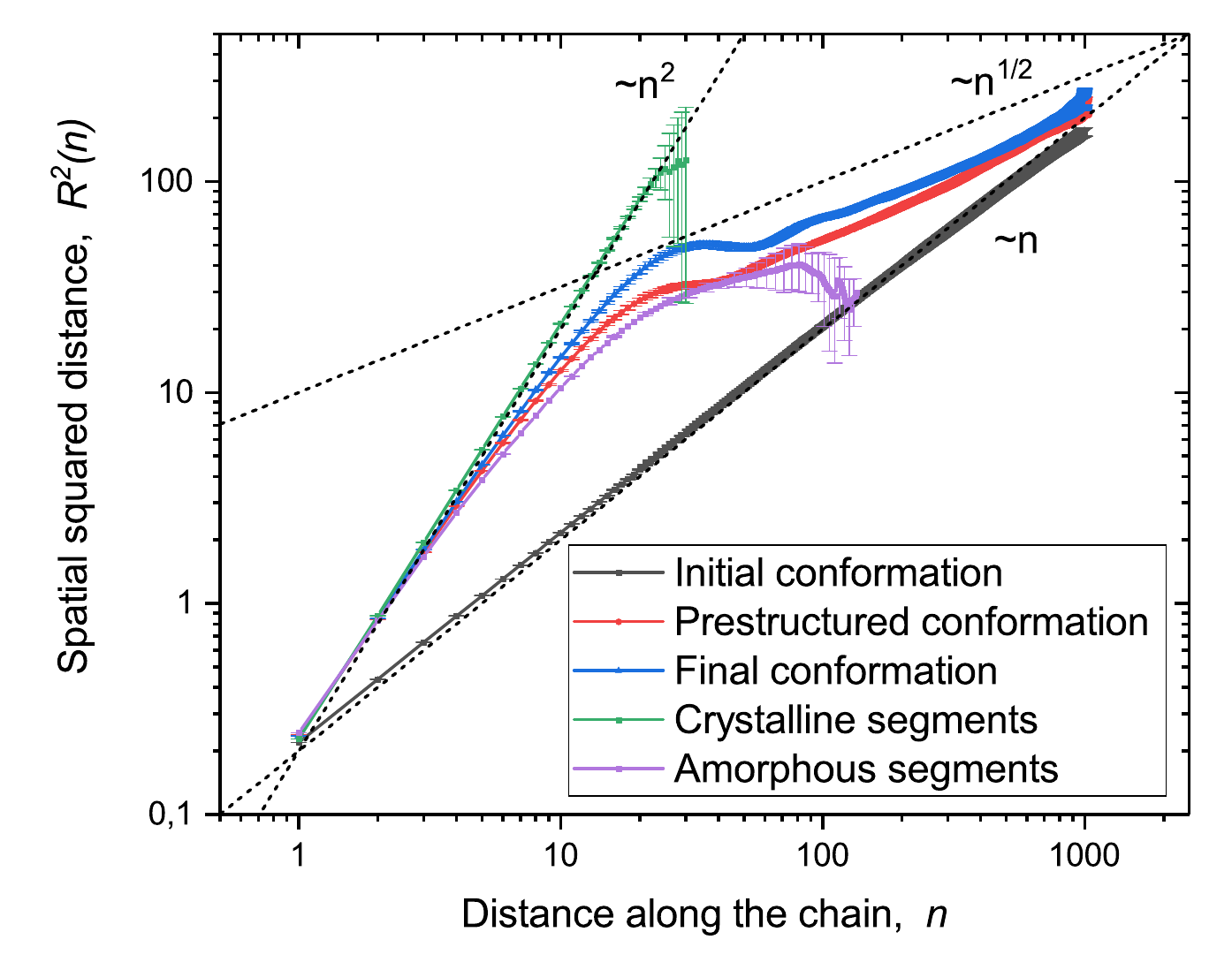}\\
		b) \includegraphics[width=0.95\linewidth]{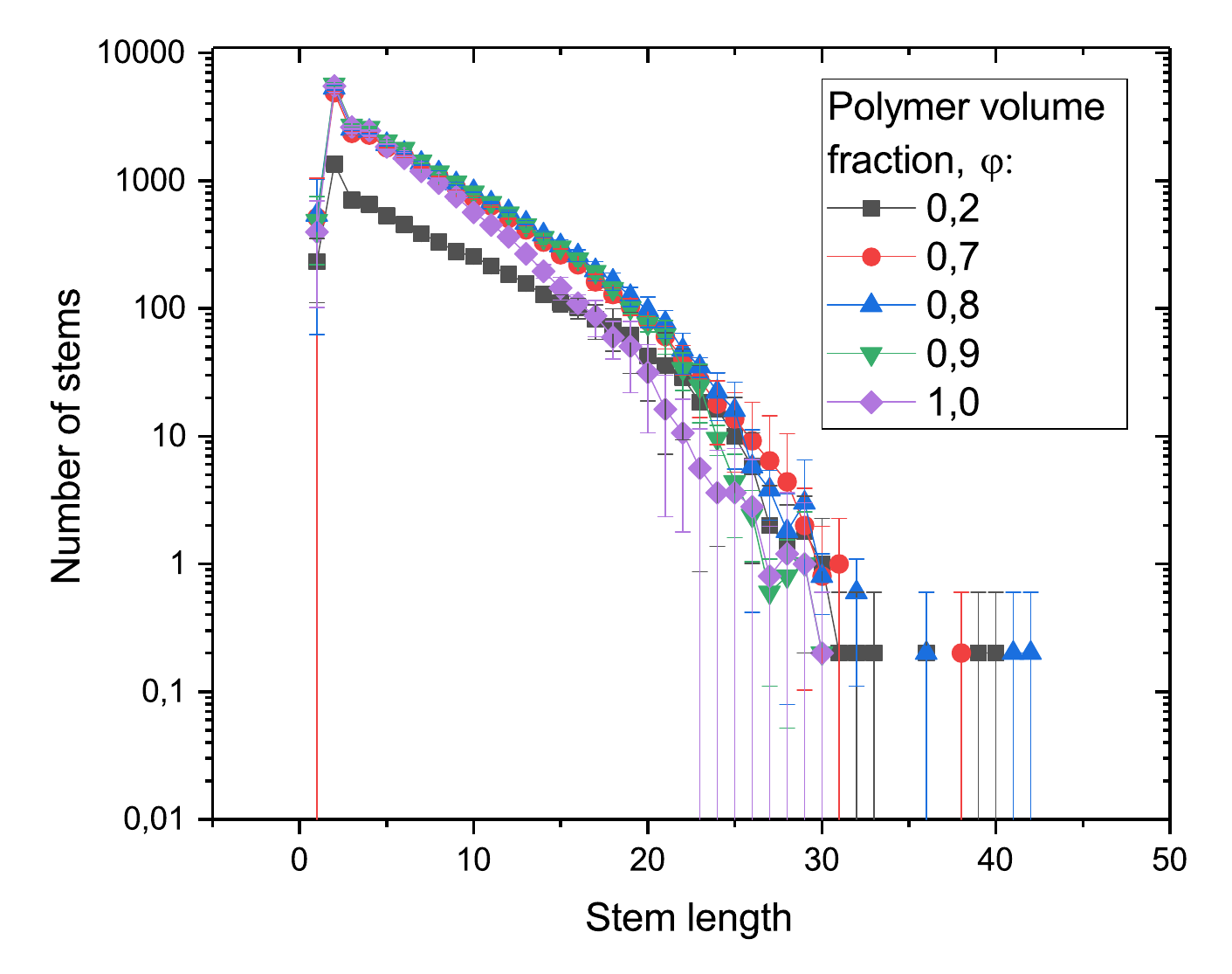}
		\caption{ Squared spatial distance $R^2$ between two beads versus distance $n$ between them along the chain for initial (blue line) and final (green curve) conformations of a system with polymer volume fraction $\varphi=0.9$ \textbf{(a)}. For the final conformation, dependencies $R^2(n)$ are plotted also separately for crystalline (black line) and amorphous (red curve) segments \textbf{(a)}. Distribution of length of crystalline segments (stems) in systems with different polymer volume fraction (shown in the legend) at the end of simulations at time $10^8$ DPD steps \textbf{(b)}. }
		\label{r2_new}
\end{figure}	

	In order to understand the structure of crystallites we have studied the conformations of chain segments inside and outside crystallites. In  Fig. \ref{r2_new}a we plot the average squared spatial distance $R^2$ between two monomer units versus the distance $n$ between them along the chain for the system configurations in the beginning, after preliminary structuring and at the end of simulation. We present the data for the system with $\varphi=0.9$ only, but actually the systems with other values of polymer volume fraction have similar behavior. The initial starting conformation of all chains in the system was Gaussian, so that the corresponding dependence lies exactly along the line $R^2(n) = l^2 n$ (black curve), where $l$ - size of the monomer unit. The data for the preliminary structured (red curve) are also shown. Finally, the data for the final time point of simulation are shown both for whole chains (green curve) as well as separately for crystalline (green curve) and amorphous (violet curve) segments. It is clear that the crystalline stems have rod-like conformations which leads to the power law dependence $R^2(n) \sim n^2$. Contrary to that, amorphous segments have quite extended conformations only on small length scales along the chain, and on larger scales they smoothly turn to much more compact conformations. Note, that the crystalline segments have quite short length (not larger than 30 beads) while the amorphous segments can reach the length of about 200 beads. The curve for whole chains, after the steep increasing $R^2 \sim n^2$, shows a plateau starting at $n \sim 30$ (which is approximately the stem length) and ending at $n \sim 60$ (which is approximately two stem lengths), while the scaling on large distances along the chain ($n > 60 \approx N_e$) is $R^2 \sim n^{1/2}$, i.e., $R \sim n^{1/4}$. We again note here that the average $N_e$ values do not change during crystallization process (see Fig.S2 in Supplementary Material). At the scale of the whole chain ($n \approx 1000$) the distance between beads does not change too much if one compares the initial Gaussian conformation before crystallization, $R^2(1000) \approx 180$, and the final one, $R^2(1000) \approx 250$, and this observation is in a good qualitative agreement with experimental data from SANS~\cite{Sadler77, Fischer1984, Fischer1988} and NMR~\cite{HongPRL2015, HongMacromol2015, HongACSLett2016}. The observed $R \sim n^{1/4}$ scaling means effectively ``more strongly crumpled'' conformations at the scales larger than the stem segment size in a crystallite, in comparison to the scaling for a single usual crumpled (fractal) globule~\cite{Nechaev}. Scaling $R \sim n^{1/4}$ was first reported for lattice animals~\cite{KhokhlovNechaev} and can also mean effectively more dense packing of some interpenetrating objects like soft spheres or Gaussian blobs. In our systems we have the case of more dense packing of hairpins (or a loose lamellae consisting of a few stems) from different chains (or from the same chain but separated from each other along the chain) in a single dense crystalline lamella (i.e., hairpins could be packed more densely than their linear size assumes). 

\begin{figure}[htbp]
        \includegraphics[width=\linewidth]{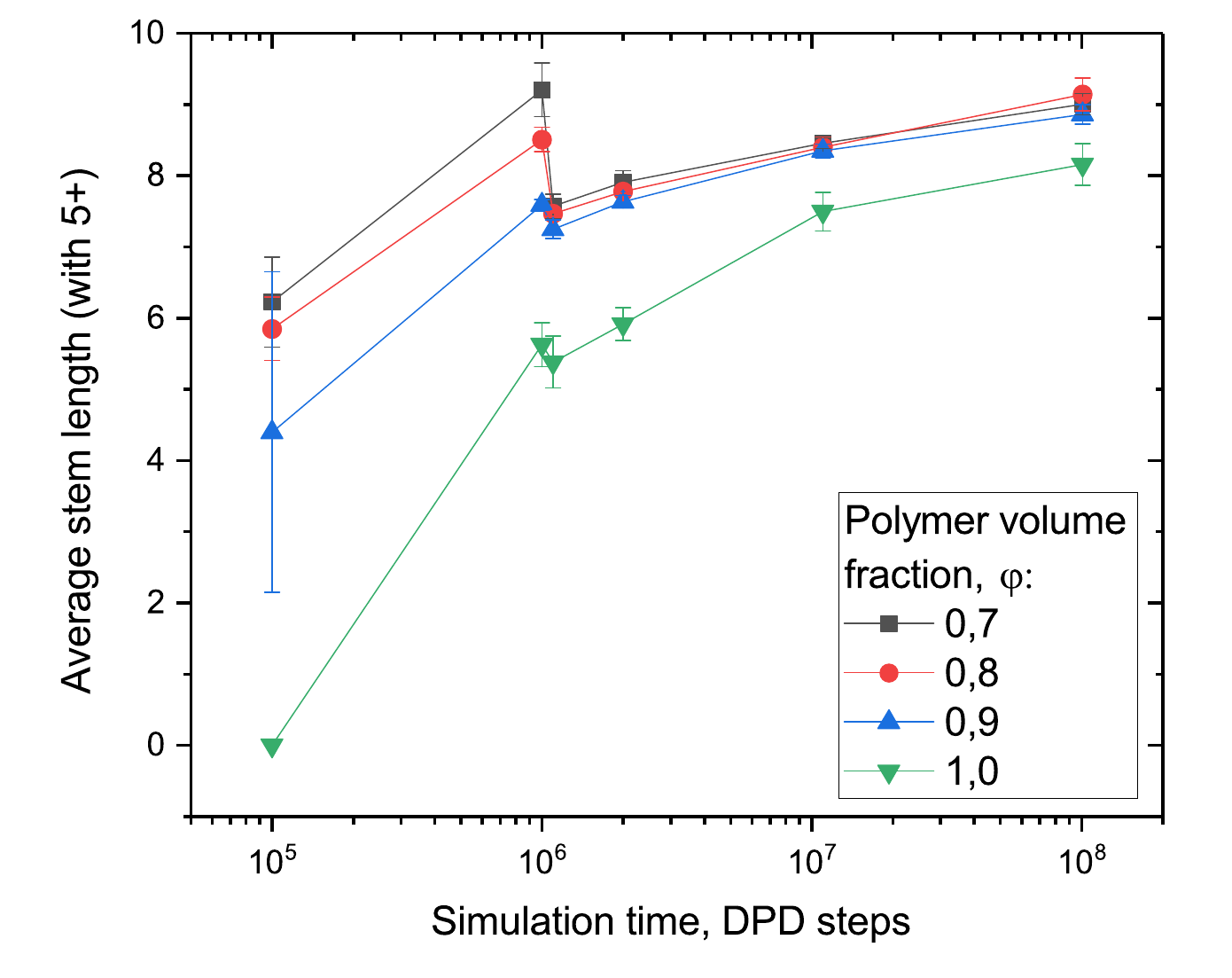}
		\caption{ Dependence of the average stem length on time for systems with different polymer volume fraction (shown in the legend). }
		\label{stem-length-time}
	\end{figure}	

	The final distributions of the length of crystallized segments for different polymer volume fractions are presented in Figure~\ref{r2_new}b. We observe the exponential Flory-like distribution of segment lengths, and these curves look very similar for all studied systems, with exception of $\varphi=1.0$ where structuring processes slowed down because of the absence of solvent. We can conclude from this plot that the internal structure of all crystallites is actually the same for all systems under study, with approximately the same maximal lamella thickness around 25-30 monomer units.
	
	In Fig.~\ref{stem-length-time} we plot the time dependence of the average stem length on large times. The average was calculated over the distribution shown in Fig.~\ref{r2_new}b and over the similar distributions calculated at time moments $10^6$ and $10^7$ DPD steps. One can see sharp jump in the dependency of average stem length at $10^6$ DPD steps. This is due to the instantaneous changing from $\theta$ to poor solvent. For the averaging over distributions we have chosen the stem size equal to 5 as the left bound of the averaging interval. The result in Fig.\ref{stem-length-time} resembles the logarithmic in time crystalline lamellar thickening~\cite{mandelkern2002crystallization, mandelkern2004crystallization, SperlingBook, StroblBook, ReiterStrobl, muthukumar2004review, Crystals2017}.

	\section{\label{conclusions}Conclusions}
	
	In this study, we have performed coarse-grained DPD computer simulations of crystallization process of long semiflexible polymer chains (of the length $N=1000$ monomer units) in melts and solutions with various polymer volume fractions and under poor solvent conditions. 
	Our DPD model appropriately modified by a (tangent-hard-spheres-like) stiffness potential, which is suitable to study general aspects of crystallization in polymers. We concentrate only on a single aspect -- crystallization of semiflexible polymers with simultaneous polymer-solvent separation after preliminary structuring in $\theta$-solvent. It is typical for many polymer processing schemes~\cite{eichhorn2009handbook}. For this particular process we try to reveal some general features which do not depend on chemical structure but are due to universal polymer properties like chain connectivity, intrachain stiffness and topology/entanglements. We start with a randomly prepared homogeneous configuration and monitor the crystallization process for a reasonably long time reaching the later stages of crystallization like, e.g., logarithmic in time lamellar thickening. We present dependencies of several observable values on the polymer volume fraction $\varphi$ because this parameter governs the role of entanglements, e.g., lamellar thickness depends on $\varphi$. Parameters of the model are chosen in such a way that the pure polymeric melt ($\varphi=1$) crystallizes (due to the steric interactions and intrachain stiffness). In solutions ($\varphi<1$), the polymer-solvent separation on local scales (demixing) occurs very fast, and this induces subsequent crystallization (like it was also observed in similar models for very dilute solutions~\cite{Wang2015}).
	 
	We have observed the following general features of crystallization behavior in our model systems:
    
	\begin{enumerate}
		\item [1.]	We observe two-stage scenario of crystallization in systems with different polymer volume fraction $\varphi$ in our model. In the first stage, the precursors of crystallites (seeds) are formed. At the end of this stage, the number of crystallites reaches its maximal value, and a steep increasing of the degree of crystallinity is observed. In the second stage, the initial crystallites grow and merge into larger crystallites, and this is observed as much more slow steady-state crystallization process. This process on large times is characterized by the same logarithmic time dependence of the degree of crystallinity for polymer volume fractions $\varphi>0.5$.
		
		\item [2.] The overall degree of crystallinity at the end of simulation decreases with increasing the polymer volume fraction in the system.
        
		\item [3.]	In our model, there is always a Flory-like length distribution of crystalline stems. The crystalline lamella thickness is calculated as an average over stem length distribution. A lamella is consisting of many rod-like stems with the size about 20-30 monomer units, and there is a hexagonal packing of stems in a lamella. We observe the logarithmic in time lamellar thickening. Each crystallite (lamella) can consist of different chains and each chain can participate in several crystallites (lamellae).
        
		\item [4.]	There is a non-monotonous dependence of the average crystallite size on the polymer volume fraction $\varphi$. We suppose that $\varphi \in [0.7, 0.95]$ is the optimal polymer volume fraction in a sense of balance between available polymer material to build crystallites and chain entanglements preventing from crystallites growth and merging.
	\end{enumerate}

	Our results are in agreement with intramolecular nucleation model of Hu et.al.~\cite{HuFrenkel}. Although our model is a coarse--grained one, it can reflect some general features of crystallization mechanisms in solutions and melts of semiflexible polymers. Moreover, our results are in qualitative agreement with Muthukumar's results on ``baby nuclei'' appearing and merging~\cite{Muthukumar2003}, although we have not checked this quantitatively. We observe both the primary and secondary nucleation (growth of crystalline nuclei) in our model as well as merging of crystallites due to filling the gap between them after intermediate amorphous regions became crystalline. Primary nucleation happens on very short times, we have not studied it in detail and it will be the topic of further research.

\begin{acknowledgements}

The research is carried out using the equipment of the shared research facilities of HPC computing resources at Lomonosov Moscow State University~\cite{voevodin2019lomonosov} with financial support of Russian Foundation for Basic Research (project 18-03-01087). PIK thanks the Hanns-Seidel-Stiftung for the financial support. We thank A.Gavrilov for providing us with DPD computer code and for multiple fruitful discussions.

\end{acknowledgements}

\bibliography{mybib}

\begin{thebibliography}{88}%
\makeatletter
\providecommand \@ifxundefined [1]{%
 \@ifx{#1\undefined}
}%
\providecommand \@ifnum [1]{%
 \ifnum #1\expandafter \@firstoftwo
 \else \expandafter \@secondoftwo
 \fi
}%
\providecommand \@ifx [1]{%
 \ifx #1\expandafter \@firstoftwo
 \else \expandafter \@secondoftwo
 \fi
}%
\providecommand \natexlab [1]{#1}%
\providecommand \enquote  [1]{``#1''}%
\providecommand \bibnamefont  [1]{#1}%
\providecommand \bibfnamefont [1]{#1}%
\providecommand \citenamefont [1]{#1}%
\providecommand \href@noop [0]{\@secondoftwo}%
\providecommand \href [0]{\begingroup \@sanitize@url \@href}%
\providecommand \@href[1]{\@@startlink{#1}\@@href}%
\providecommand \@@href[1]{\endgroup#1\@@endlink}%
\providecommand \@sanitize@url [0]{\catcode `\\12\catcode `\$12\catcode
  `\&12\catcode `\#12\catcode `\^12\catcode `\_12\catcode `\%12\relax}%
\providecommand \@@startlink[1]{}%
\providecommand \@@endlink[0]{}%
\providecommand \url  [0]{\begingroup\@sanitize@url \@url }%
\providecommand \@url [1]{\endgroup\@href {#1}{\urlprefix }}%
\providecommand \urlprefix  [0]{URL }%
\providecommand \Eprint [0]{\href }%
\providecommand \doibase [0]{http://dx.doi.org/}%
\providecommand \selectlanguage [0]{\@gobble}%
\providecommand \bibinfo  [0]{\@secondoftwo}%
\providecommand \bibfield  [0]{\@secondoftwo}%
\providecommand \translation [1]{[#1]}%
\providecommand \BibitemOpen [0]{}%
\providecommand \bibitemStop [0]{}%
\providecommand \bibitemNoStop [0]{.\EOS\space}%
\providecommand \EOS [0]{\spacefactor3000\relax}%
\providecommand \BibitemShut  [1]{\csname bibitem#1\endcsname}%
\let\auto@bib@innerbib\@empty
\bibitem [{\citenamefont {Mandelkern}(2002)}]{mandelkern2002crystallization}%
  \BibitemOpen
  \bibfield  {author} {\bibinfo {author} {\bibfnamefont {L.}~\bibnamefont
  {Mandelkern}},\ }\href@noop {} {\emph {\bibinfo {title} {Crystallization of
  Polymers: Volume 1, Equilibrium Concepts}}}\ (\bibinfo  {publisher}
  {Cambridge University Press},\ \bibinfo {year} {2002})\BibitemShut {NoStop}%
\bibitem [{\citenamefont {Mandelkern}(2004)}]{mandelkern2004crystallization}%
  \BibitemOpen
  \bibfield  {author} {\bibinfo {author} {\bibfnamefont {L.}~\bibnamefont
  {Mandelkern}},\ }\href@noop {} {\emph {\bibinfo {title} {Crystallization of
  Polymers: Volume 2, Kinetics and Mechanisms}}}\ (\bibinfo  {publisher}
  {Cambridge University Press},\ \bibinfo {year} {2004})\BibitemShut {NoStop}%
\bibitem [{\citenamefont {Sperling}(2006)}]{SperlingBook}%
  \BibitemOpen
  \bibfield  {author} {\bibinfo {author} {\bibfnamefont {L.~H.}\ \bibnamefont
  {Sperling}},\ }\href@noop {} {\emph {\bibinfo {title} {Introduction to
  Physical Polymer Science, 4-th Edition}}}\ (\bibinfo  {publisher} {John Wiley
  \& Sons, Inc.},\ \bibinfo {year} {2006})\BibitemShut {NoStop}%
\bibitem [{\citenamefont {Strobl}(2007)}]{StroblBook}%
  \BibitemOpen
  \bibfield  {author} {\bibinfo {author} {\bibfnamefont {G.~R.}\ \bibnamefont
  {Strobl}},\ }\href@noop {} {\emph {\bibinfo {title} {The Physics of Polymers:
  Concepts for Understanding Their Structures and Behavior, 3-rd Edition}}}\
  (\bibinfo  {publisher} {Springer-Verlag Berlin Heidelberg},\ \bibinfo {year}
  {2007})\BibitemShut {NoStop}%
\bibitem [{\citenamefont {Reiter}\ and\ \citenamefont
  {Strobl}(2007)}]{ReiterStrobl}%
  \BibitemOpen
  \bibinfo {editor} {\bibfnamefont {G.}~\bibnamefont {Reiter}}\ and\ \bibinfo
  {editor} {\bibfnamefont {G.~R.}\ \bibnamefont {Strobl}},\ eds.,\ \href@noop
  {} {\emph {\bibinfo {title} {Progress in Understanding of Polymer
  Crystallization}}},\ Lect. Notes Phys. 714\ (\bibinfo  {publisher}
  {Springer-Verlag},\ \bibinfo {address} {Berlin Heidelberg},\ \bibinfo {year}
  {2007})\BibitemShut {NoStop}%
\bibitem [{\citenamefont {Muthukumar}(2004)}]{muthukumar2004review}%
  \BibitemOpen
  \bibfield  {author} {\bibinfo {author} {\bibfnamefont {M.}~\bibnamefont
  {Muthukumar}},\ }\bibfield  {title} {\enquote {\bibinfo {title} {Nucleation
  in polymer crystallization},}\ }\href@noop {} {\bibfield  {journal} {\bibinfo
   {journal} {Advances in Chemical Physics}\ }\textbf {\bibinfo {volume}
  {128}},\ \bibinfo {pages} {1--64} (\bibinfo {year} {2004})}\BibitemShut
  {NoStop}%
\bibitem [{\citenamefont {Zhang}, \citenamefont {Guo},\ and\ \citenamefont
  {Xu}(2017)}]{Crystals2017}%
  \BibitemOpen
  \bibfield  {author} {\bibinfo {author} {\bibfnamefont {M.~C.}\ \bibnamefont
  {Zhang}}, \bibinfo {author} {\bibfnamefont {B.-H.}\ \bibnamefont {Guo}}, \
  and\ \bibinfo {author} {\bibfnamefont {J.}~\bibnamefont {Xu}},\ }\bibfield
  {title} {\enquote {\bibinfo {title} {A review on polymer crystallization
  theories},}\ }\href {\doibase 10.3390/cryst7010004} {\bibfield  {journal}
  {\bibinfo  {journal} {Crystals}\ }\textbf {\bibinfo {volume} {7}},\ \bibinfo
  {pages} {4} (\bibinfo {year} {2017})}\BibitemShut {NoStop}%
\bibitem [{\citenamefont {Doi}\ and\ \citenamefont
  {Edwards}(1988)}]{DoiEdwards}%
  \BibitemOpen
  \bibfield  {author} {\bibinfo {author} {\bibfnamefont {M.}~\bibnamefont
  {Doi}}\ and\ \bibinfo {author} {\bibfnamefont {S.~F.}\ \bibnamefont
  {Edwards}},\ }\href@noop {} {\emph {\bibinfo {title} {The Theory of Polymer
  Dynamics}}}\ (\bibinfo  {publisher} {Clarendon Press, Oxford},\ \bibinfo
  {year} {1988})\BibitemShut {NoStop}%
\bibitem [{\citenamefont {Allegra}(1977)}]{Allegra1977}%
  \BibitemOpen
  \bibfield  {author} {\bibinfo {author} {\bibfnamefont {G.}~\bibnamefont
  {Allegra}},\ }\bibfield  {title} {\enquote {\bibinfo {title} {Chain folding
  and polymer crystallization: A statistical--mechanical approach},}\ }\href
  {\doibase 10.1063/1.433919} {\bibfield  {journal} {\bibinfo  {journal} {The
  Journal of Chemical Physics}\ }\textbf {\bibinfo {volume} {66}},\ \bibinfo
  {pages} {5453--5463} (\bibinfo {year} {1977})}\BibitemShut {NoStop}%
\bibitem [{\citenamefont {Muthukumar}(2003)}]{Muthukumar2003}%
  \BibitemOpen
  \bibfield  {author} {\bibinfo {author} {\bibfnamefont {M.}~\bibnamefont
  {Muthukumar}},\ }\bibfield  {title} {\enquote {\bibinfo {title} {{Molecular
  modelling of nucleation in polymers}},}\ }\href {\doibase
  10.1098/rsta.2002.1149} {\bibfield  {journal} {\bibinfo  {journal}
  {Philosophical Transactions of the Royal Society A: Mathematical, Physical
  and Engineering Sciences}\ }\textbf {\bibinfo {volume} {361}},\ \bibinfo
  {pages} {539--556} (\bibinfo {year} {2003})}\BibitemShut {NoStop}%
\bibitem [{\citenamefont {Allegra}\ and\ \citenamefont
  {Famulari}(2009)}]{Allegra2009}%
  \BibitemOpen
  \bibfield  {author} {\bibinfo {author} {\bibfnamefont {G.}~\bibnamefont
  {Allegra}}\ and\ \bibinfo {author} {\bibfnamefont {A.}~\bibnamefont
  {Famulari}},\ }\bibfield  {title} {\enquote {\bibinfo {title} {Chain
  statistics in polyethylene crystallization},}\ }\href {\doibase
  10.1016/j.polymer.2009.01.063} {\bibfield  {journal} {\bibinfo  {journal}
  {Polymer}\ }\textbf {\bibinfo {volume} {50}},\ \bibinfo {pages} {1819--1829}
  (\bibinfo {year} {2009})}\BibitemShut {NoStop}%
\bibitem [{\citenamefont {Stepanow}(2014)}]{Stepanow2014}%
  \BibitemOpen
  \bibfield  {author} {\bibinfo {author} {\bibfnamefont {S.}~\bibnamefont
  {Stepanow}},\ }\bibfield  {title} {\enquote {\bibinfo {title} {{Kinetic
  mechanism of chain folding in polymer crystallization}},}\ }\href {\doibase
  10.1103/PhysRevE.90.032601} {\bibfield  {journal} {\bibinfo  {journal}
  {Physical Review E}\ }\textbf {\bibinfo {volume} {90}},\ \bibinfo {pages}
  {1--14} (\bibinfo {year} {2014})}\BibitemShut {NoStop}%
\bibitem [{\citenamefont {Strobl}(2000)}]{Strobl2000}%
  \BibitemOpen
  \bibfield  {author} {\bibinfo {author} {\bibfnamefont {G.}~\bibnamefont
  {Strobl}},\ }\bibfield  {title} {\enquote {\bibinfo {title} {From the melt
  via mesomorphic and granular crystalline layers to lamellar crystallites: A
  major route followed in polymer crystallization?}}\ }\href {\doibase
  10.1007/s101890070030} {\bibfield  {journal} {\bibinfo  {journal} {The
  European Physical Journal E}\ }\textbf {\bibinfo {volume} {3}},\ \bibinfo
  {pages} {165--183} (\bibinfo {year} {2000})}\BibitemShut {NoStop}%
\bibitem [{\citenamefont {Strobl}(2006)}]{Strobl2006}%
  \BibitemOpen
  \bibfield  {author} {\bibinfo {author} {\bibfnamefont {G.}~\bibnamefont
  {Strobl}},\ }\bibfield  {title} {\enquote {\bibinfo {title} {{Crystallization
  and melting of bulk polymers: New observations, conclusions and a
  thermodynamic scheme}},}\ }\href {\doibase
  10.1016/j.progpolymsci.2006.01.001} {\bibfield  {journal} {\bibinfo
  {journal} {Progress in Polymer Science (Oxford)}\ }\textbf {\bibinfo {volume}
  {31}},\ \bibinfo {pages} {398--442} (\bibinfo {year} {2006})}\BibitemShut
  {NoStop}%
\bibitem [{\citenamefont {Strobl}(2009)}]{Strobl2009}%
  \BibitemOpen
  \bibfield  {author} {\bibinfo {author} {\bibfnamefont {G.}~\bibnamefont
  {Strobl}},\ }\bibfield  {title} {\enquote {\bibinfo {title} {Colloquium: Laws
  controlling crystallization and melting in bulk polymers},}\ }\href {\doibase
  10.1103/RevModPhys.81.1287} {\bibfield  {journal} {\bibinfo  {journal}
  {Reviews of Modern Physics}\ }\textbf {\bibinfo {volume} {81}},\ \bibinfo
  {pages} {1287--1300} (\bibinfo {year} {2009})}\BibitemShut {NoStop}%
\bibitem [{\citenamefont {Kundagrami}\ and\ \citenamefont
  {Muthukumar}(2007)}]{KundMuthu2007}%
  \BibitemOpen
  \bibfield  {author} {\bibinfo {author} {\bibfnamefont {A.}~\bibnamefont
  {Kundagrami}}\ and\ \bibinfo {author} {\bibfnamefont {M.}~\bibnamefont
  {Muthukumar}},\ }\bibfield  {title} {\enquote {\bibinfo {title} {Continuum
  theory of polymer crystallization},}\ }\href@noop {} {\bibfield  {journal}
  {\bibinfo  {journal} {The Journal of Chemical Physics}\ }\textbf {\bibinfo
  {volume} {126}},\ \bibinfo {pages} {144901} (\bibinfo {year}
  {2007})}\BibitemShut {NoStop}%
\bibitem [{\citenamefont {Allegra}\ and\ \citenamefont
  {Meille}(1999)}]{Allegra1999}%
  \BibitemOpen
  \bibfield  {author} {\bibinfo {author} {\bibfnamefont {G.}~\bibnamefont
  {Allegra}}\ and\ \bibinfo {author} {\bibfnamefont {S.}~\bibnamefont
  {Meille}},\ }\bibfield  {title} {\enquote {\bibinfo {title} {The bundle
  theory for polymer crystallisation},}\ }\href {\doibase 10.1039/A905658K}
  {\bibfield  {journal} {\bibinfo  {journal} {Physical Chemistry Chemical
  Physics}\ }\textbf {\bibinfo {volume} {1}},\ \bibinfo {pages} {5179--5188}
  (\bibinfo {year} {1999})}\BibitemShut {NoStop}%
\bibitem [{\citenamefont {Luo}\ and\ \citenamefont {Sommer}(2011)}]{Luo2011}%
  \BibitemOpen
  \bibfield  {author} {\bibinfo {author} {\bibfnamefont {C.}~\bibnamefont
  {Luo}}\ and\ \bibinfo {author} {\bibfnamefont {J.~U.}\ \bibnamefont
  {Sommer}},\ }\bibfield  {title} {\enquote {\bibinfo {title} {{Growth pathway
  and precursor states in single lamellar crystallization: MD simulations}},}\
  }\href {\doibase 10.1021/ma102380m} {\bibfield  {journal} {\bibinfo
  {journal} {Macromolecules}\ }\textbf {\bibinfo {volume} {44}},\ \bibinfo
  {pages} {1523--1529} (\bibinfo {year} {2011})}\BibitemShut {NoStop}%
\bibitem [{\citenamefont {Muthukumar}(2000)}]{Muthukumar2000}%
  \BibitemOpen
  \bibfield  {author} {\bibinfo {author} {\bibfnamefont {M.}~\bibnamefont
  {Muthukumar}},\ }\bibfield  {title} {\enquote {\bibinfo {title} {{Commentary
  on theories of polymer crystallization}},}\ }\href {\doibase
  10.1007/s101890070033} {\bibfield  {journal} {\bibinfo  {journal} {The
  European Physical Journal E}\ }\textbf {\bibinfo {volume} {3}},\ \bibinfo
  {pages} {199--202} (\bibinfo {year} {2000})}\BibitemShut {NoStop}%
\bibitem [{\citenamefont {Gee}, \citenamefont {Lacevic},\ and\ \citenamefont
  {Fried}(2006)}]{gee2006atomistic}%
  \BibitemOpen
  \bibfield  {author} {\bibinfo {author} {\bibfnamefont {R.~H.}\ \bibnamefont
  {Gee}}, \bibinfo {author} {\bibfnamefont {N.}~\bibnamefont {Lacevic}}, \ and\
  \bibinfo {author} {\bibfnamefont {L.~E.}\ \bibnamefont {Fried}},\ }\bibfield
  {title} {\enquote {\bibinfo {title} {Atomistic simulations of spinodal phase
  separation preceding polymer crystallization},}\ }\href@noop {} {\bibfield
  {journal} {\bibinfo  {journal} {Nature materials}\ }\textbf {\bibinfo
  {volume} {5}},\ \bibinfo {pages} {39--43} (\bibinfo {year}
  {2006})}\BibitemShut {NoStop}%
\bibitem [{\citenamefont {Elliott}(2011)}]{Elliott2011}%
  \BibitemOpen
  \bibfield  {author} {\bibinfo {author} {\bibfnamefont {J.~A.}\ \bibnamefont
  {Elliott}},\ }\bibfield  {title} {\enquote {\bibinfo {title} {Novel
  approaches to multiscale modelling in materials science},}\ }\href@noop {}
  {\bibfield  {journal} {\bibinfo  {journal} {International Materials Reviews}\
  }\textbf {\bibinfo {volume} {56}},\ \bibinfo {pages} {207--225} (\bibinfo
  {year} {2011})}\BibitemShut {NoStop}%
\bibitem [{\citenamefont {Gooneie}, \citenamefont {Schuschnigg},\ and\
  \citenamefont {Holzer}(2017)}]{Gooneie-Polymers2017}%
  \BibitemOpen
  \bibfield  {author} {\bibinfo {author} {\bibfnamefont {A.}~\bibnamefont
  {Gooneie}}, \bibinfo {author} {\bibfnamefont {S.}~\bibnamefont
  {Schuschnigg}}, \ and\ \bibinfo {author} {\bibfnamefont {C.}~\bibnamefont
  {Holzer}},\ }\bibfield  {title} {\enquote {\bibinfo {title} {A review of
  multiscale computational methods in polymeric materials},}\ }\href {\doibase
  10.3390/polym9010016} {\bibfield  {journal} {\bibinfo  {journal} {Polymers}\
  }\textbf {\bibinfo {volume} {9}},\ \bibinfo {pages} {16} (\bibinfo {year}
  {2017})}\BibitemShut {NoStop}%
\bibitem [{\citenamefont {Paul}, \citenamefont {Yoon},\ and\ \citenamefont
  {Smith}(1995)}]{paul1995optimized}%
  \BibitemOpen
  \bibfield  {author} {\bibinfo {author} {\bibfnamefont {W.}~\bibnamefont
  {Paul}}, \bibinfo {author} {\bibfnamefont {D.~Y.}\ \bibnamefont {Yoon}}, \
  and\ \bibinfo {author} {\bibfnamefont {G.~D.}\ \bibnamefont {Smith}},\
  }\bibfield  {title} {\enquote {\bibinfo {title} {An optimized united atom
  model for simulations of polymethylene melts},}\ }\href@noop {} {\bibfield
  {journal} {\bibinfo  {journal} {The Journal of Chemical Physics}\ }\textbf
  {\bibinfo {volume} {103}},\ \bibinfo {pages} {1702--1709} (\bibinfo {year}
  {1995})}\BibitemShut {NoStop}%
\bibitem [{\citenamefont {Yamamoto}(1997)}]{Yamamoto1997}%
  \BibitemOpen
  \bibfield  {author} {\bibinfo {author} {\bibfnamefont {T.}~\bibnamefont
  {Yamamoto}},\ }\bibfield  {title} {\enquote {\bibinfo {title} {Molecular
  dynamics simulation of polymer crystallization through chain folding},}\
  }\href@noop {} {\bibfield  {journal} {\bibinfo  {journal} {The Journal of
  Chemical Physics}\ }\textbf {\bibinfo {volume} {107}},\ \bibinfo {pages}
  {2653--2663} (\bibinfo {year} {1997})}\BibitemShut {NoStop}%
\bibitem [{\citenamefont {Yamamoto}(1998)}]{Yamamoto1998I}%
  \BibitemOpen
  \bibfield  {author} {\bibinfo {author} {\bibfnamefont {T.}~\bibnamefont
  {Yamamoto}},\ }\bibfield  {title} {\enquote {\bibinfo {title}
  {Molecular-dynamics simulation of polymer ordering. i. crystallization from
  vapor phase},}\ }\href@noop {} {\bibfield  {journal} {\bibinfo  {journal}
  {The Journal of Chemical Physics}\ }\textbf {\bibinfo {volume} {109}},\
  \bibinfo {pages} {4638--4645} (\bibinfo {year} {1998})}\BibitemShut {NoStop}%
\bibitem [{\citenamefont {Waheed}, \citenamefont {Lavine},\ and\ \citenamefont
  {Rutledge}(2002)}]{Rutledge2002}%
  \BibitemOpen
  \bibfield  {author} {\bibinfo {author} {\bibfnamefont {N.}~\bibnamefont
  {Waheed}}, \bibinfo {author} {\bibfnamefont {M.~S.}\ \bibnamefont {Lavine}},
  \ and\ \bibinfo {author} {\bibfnamefont {G.~C.}\ \bibnamefont {Rutledge}},\
  }\bibfield  {title} {\enquote {\bibinfo {title} {Molecular simulation of
  crystal growth in n- eicosane},}\ }\href@noop {} {\bibfield  {journal}
  {\bibinfo  {journal} {The Journal of Chemical Physics}\ }\textbf {\bibinfo
  {volume} {116}},\ \bibinfo {pages} {2301--2309} (\bibinfo {year}
  {2002})}\BibitemShut {NoStop}%
\bibitem [{\citenamefont {Ko}\ \emph {et~al.}(2004)\citenamefont {Ko},
  \citenamefont {Waheed}, \citenamefont {Lavine},\ and\ \citenamefont
  {Rutledge}}]{Rutledge2004}%
  \BibitemOpen
  \bibfield  {author} {\bibinfo {author} {\bibfnamefont {M.~J.}\ \bibnamefont
  {Ko}}, \bibinfo {author} {\bibfnamefont {N.}~\bibnamefont {Waheed}}, \bibinfo
  {author} {\bibfnamefont {M.~S.}\ \bibnamefont {Lavine}}, \ and\ \bibinfo
  {author} {\bibfnamefont {G.~C.}\ \bibnamefont {Rutledge}},\ }\bibfield
  {title} {\enquote {\bibinfo {title} {Characterization of polyethylene
  crystallization from an oriented melt by molecular dynamics simulation},}\
  }\href@noop {} {\bibfield  {journal} {\bibinfo  {journal} {The Journal of
  Chemical Physics}\ }\textbf {\bibinfo {volume} {121}},\ \bibinfo {pages}
  {2823--2832} (\bibinfo {year} {2004})}\BibitemShut {NoStop}%
\bibitem [{\citenamefont {Meyer}\ and\ \citenamefont
  {M{\"{u}}ller-Plathe}(2001)}]{Meyer2001}%
  \BibitemOpen
  \bibfield  {author} {\bibinfo {author} {\bibfnamefont {H.}~\bibnamefont
  {Meyer}}\ and\ \bibinfo {author} {\bibfnamefont {F.}~\bibnamefont
  {M{\"{u}}ller-Plathe}},\ }\bibfield  {title} {\enquote {\bibinfo {title}
  {{Formation of chain-folded structures in supercooled polymer melts}},}\
  }\href {\doibase 10.1063/1.1415456} {\bibfield  {journal} {\bibinfo
  {journal} {The Journal of Chemical Physics}\ }\textbf {\bibinfo {volume}
  {115}},\ \bibinfo {pages} {7807--7810} (\bibinfo {year} {2001})}\BibitemShut
  {NoStop}%
\bibitem [{\citenamefont {Meyer}\ and\ \citenamefont
  {M{\"{u}}ller-Plathe}(2002)}]{Meyer2002}%
  \BibitemOpen
  \bibfield  {author} {\bibinfo {author} {\bibfnamefont {H.}~\bibnamefont
  {Meyer}}\ and\ \bibinfo {author} {\bibfnamefont {F.}~\bibnamefont
  {M{\"{u}}ller-Plathe}},\ }\bibfield  {title} {\enquote {\bibinfo {title}
  {{Formation of chain-folded structures in supercooled polymer melts examined
  by MD simulations}},}\ }\href {\doibase 10.1021/ma011309l} {\bibfield
  {journal} {\bibinfo  {journal} {Macromolecules}\ }\textbf {\bibinfo {volume}
  {35}},\ \bibinfo {pages} {1241--1252} (\bibinfo {year} {2002})}\BibitemShut
  {NoStop}%
\bibitem [{\citenamefont {Vettorel}\ and\ \citenamefont
  {Meyer}(2006)}]{vettorel2006coarse}%
  \BibitemOpen
  \bibfield  {author} {\bibinfo {author} {\bibfnamefont {T.}~\bibnamefont
  {Vettorel}}\ and\ \bibinfo {author} {\bibfnamefont {H.}~\bibnamefont
  {Meyer}},\ }\bibfield  {title} {\enquote {\bibinfo {title} {Coarse graining
  of short polythylene chains for studying polymer crystallization},}\
  }\href@noop {} {\bibfield  {journal} {\bibinfo  {journal} {Journal of
  Chemical Theory and Computation}\ }\textbf {\bibinfo {volume} {2}},\ \bibinfo
  {pages} {616--629} (\bibinfo {year} {2006})}\BibitemShut {NoStop}%
\bibitem [{\citenamefont {Vettorel}\ \emph {et~al.}(2007)\citenamefont
  {Vettorel}, \citenamefont {Meyer}, \citenamefont {Baschnagel},\ and\
  \citenamefont {Fuchs}}]{Vettorel2007}%
  \BibitemOpen
  \bibfield  {author} {\bibinfo {author} {\bibfnamefont {T.}~\bibnamefont
  {Vettorel}}, \bibinfo {author} {\bibfnamefont {H.}~\bibnamefont {Meyer}},
  \bibinfo {author} {\bibfnamefont {J.}~\bibnamefont {Baschnagel}}, \ and\
  \bibinfo {author} {\bibfnamefont {M.}~\bibnamefont {Fuchs}},\ }\bibfield
  {title} {\enquote {\bibinfo {title} {{Structural properties of crystallizable
  polymer melts: Intrachain and interchain correlation functions}},}\ }\href
  {\doibase 10.1103/PhysRevE.75.041801} {\bibfield  {journal} {\bibinfo
  {journal} {Physical Review E}\ }\textbf {\bibinfo {volume} {75}},\ \bibinfo
  {pages} {1--14} (\bibinfo {year} {2007})}\BibitemShut {NoStop}%
\bibitem [{\citenamefont {Shakirov}\ and\ \citenamefont
  {Paul}(2018)}]{Paul-Shakirov}%
  \BibitemOpen
  \bibfield  {author} {\bibinfo {author} {\bibfnamefont {T.}~\bibnamefont
  {Shakirov}}\ and\ \bibinfo {author} {\bibfnamefont {W.}~\bibnamefont
  {Paul}},\ }\bibfield  {title} {\enquote {\bibinfo {title} {Crystallization in
  melts of short, semiflexible hard polymer chains: An interplay of entropies
  and dimensions},}\ }\href@noop {} {\bibfield  {journal} {\bibinfo  {journal}
  {Physical Review E}\ }\textbf {\bibinfo {volume} {97}},\ \bibinfo {pages}
  {042501} (\bibinfo {year} {2018})}\BibitemShut {NoStop}%
\bibitem [{\citenamefont {Liu}\ and\ \citenamefont
  {Muthukumar}(1998)}]{Muthukumar1998}%
  \BibitemOpen
  \bibfield  {author} {\bibinfo {author} {\bibfnamefont {C.}~\bibnamefont
  {Liu}}\ and\ \bibinfo {author} {\bibfnamefont {M.}~\bibnamefont
  {Muthukumar}},\ }\bibfield  {title} {\enquote {\bibinfo {title} {Langevin
  dynamics simulations of early-stage polymer nucleation and
  crystallization},}\ }\href@noop {} {\bibfield  {journal} {\bibinfo  {journal}
  {The Journal of Chemical Physics}\ }\textbf {\bibinfo {volume} {109}},\
  \bibinfo {pages} {2536--2542} (\bibinfo {year} {1998})}\BibitemShut {NoStop}%
\bibitem [{\citenamefont {Welch}\ and\ \citenamefont
  {Muthukumar}(2001)}]{Welch2001}%
  \BibitemOpen
  \bibfield  {author} {\bibinfo {author} {\bibfnamefont {P.}~\bibnamefont
  {Welch}}\ and\ \bibinfo {author} {\bibfnamefont {M.}~\bibnamefont
  {Muthukumar}},\ }\bibfield  {title} {\enquote {\bibinfo {title} {{Molecular
  Mechanisms of Polymer Crystallization from Solution}},}\ }\href {\doibase
  10.1103/PhysRevLett.87.218302} {\bibfield  {journal} {\bibinfo  {journal}
  {Physical Review Letters}\ }\textbf {\bibinfo {volume} {87}},\ \bibinfo
  {pages} {218302} (\bibinfo {year} {2001})}\BibitemShut {NoStop}%
\bibitem [{\citenamefont {Muthukumar}(2005)}]{muthukumar2005modeling}%
  \BibitemOpen
  \bibfield  {author} {\bibinfo {author} {\bibfnamefont {M.}~\bibnamefont
  {Muthukumar}},\ }\bibfield  {title} {\enquote {\bibinfo {title} {Modeling
  polymer crystallization},}\ }\href@noop {} {\bibfield  {journal} {\bibinfo
  {journal} {Advances in Polymer Science}\ }\textbf {\bibinfo {volume} {191}},\
  \bibinfo {pages} {241--274} (\bibinfo {year} {2005})}\BibitemShut {NoStop}%
\bibitem [{\citenamefont {Zhang}\ and\ \citenamefont
  {Muthukumar}(2007)}]{Zhang2007}%
  \BibitemOpen
  \bibfield  {author} {\bibinfo {author} {\bibfnamefont {J.}~\bibnamefont
  {Zhang}}\ and\ \bibinfo {author} {\bibfnamefont {M.}~\bibnamefont
  {Muthukumar}},\ }\bibfield  {title} {\enquote {\bibinfo {title} {{Monte Carlo
  simulations of single crystals from polymer solutions}},}\ }\href {\doibase
  10.1063/1.2740256} {\bibfield  {journal} {\bibinfo  {journal} {The Journal of
  Chemical Physics}\ }\textbf {\bibinfo {volume} {126}},\ \bibinfo {pages}
  {234904} (\bibinfo {year} {2007})}\BibitemShut {NoStop}%
\bibitem [{\citenamefont {Welch}(2017)}]{welch2017examining}%
  \BibitemOpen
  \bibfield  {author} {\bibinfo {author} {\bibfnamefont {P.}~\bibnamefont
  {Welch}},\ }\bibfield  {title} {\enquote {\bibinfo {title} {Examining the
  role of fluctuations in the early stages of homogenous polymer
  crystallization with simulation and statistical learning},}\ }\href@noop {}
  {\bibfield  {journal} {\bibinfo  {journal} {The Journal of Chemical Physics}\
  }\textbf {\bibinfo {volume} {146}},\ \bibinfo {pages} {044901} (\bibinfo
  {year} {2017})}\BibitemShut {NoStop}%
\bibitem [{\citenamefont {Yamamoto}(2008)}]{Yamamoto2008}%
  \BibitemOpen
  \bibfield  {author} {\bibinfo {author} {\bibfnamefont {T.}~\bibnamefont
  {Yamamoto}},\ }\bibfield  {title} {\enquote {\bibinfo {title} {Molecular
  dynamics simulations of steady-state crystal growth and homogeneous
  nucleation in polyethylene-like polymer},}\ }\href@noop {} {\bibfield
  {journal} {\bibinfo  {journal} {The Journal of Chemical Physics}\ }\textbf
  {\bibinfo {volume} {129}},\ \bibinfo {pages} {184903} (\bibinfo {year}
  {2008})}\BibitemShut {NoStop}%
\bibitem [{\citenamefont {Yamamoto}(2009)}]{Yamamoto2009}%
  \BibitemOpen
  \bibfield  {author} {\bibinfo {author} {\bibfnamefont {T.}~\bibnamefont
  {Yamamoto}},\ }\bibfield  {title} {\enquote {\bibinfo {title} {{Computer
  modeling of polymer crystallization - Toward computer-assisted materials'
  design}},}\ }\href {\doibase 10.1016/j.polymer.2009.02.038} {\bibfield
  {journal} {\bibinfo  {journal} {Polymer}\ }\textbf {\bibinfo {volume} {50}},\
  \bibinfo {pages} {1975--1985} (\bibinfo {year} {2009})}\BibitemShut {NoStop}%
\bibitem [{\citenamefont {Yamamoto}(2010)}]{Yamamoto2010}%
  \BibitemOpen
  \bibfield  {author} {\bibinfo {author} {\bibfnamefont {T.}~\bibnamefont
  {Yamamoto}},\ }\bibfield  {title} {\enquote {\bibinfo {title} {Molecular
  dynamics simulations of polymer crystallization in highly supercooled melt:
  Primary nucleation and cold crystallization},}\ }\href@noop {} {\bibfield
  {journal} {\bibinfo  {journal} {The Journal of Chemical Physics}\ }\textbf
  {\bibinfo {volume} {133}},\ \bibinfo {pages} {034904} (\bibinfo {year}
  {2010})}\BibitemShut {NoStop}%
\bibitem [{\citenamefont {Yamamoto}(2013)}]{Yamamoto2013}%
  \BibitemOpen
  \bibfield  {author} {\bibinfo {author} {\bibfnamefont {T.}~\bibnamefont
  {Yamamoto}},\ }\bibfield  {title} {\enquote {\bibinfo {title} {{Molecular
  dynamics of polymer crystallization revisited: Crystallization from the melt
  and the glass in longer polyethylene}},}\ }\href@noop {} {\bibfield
  {journal} {\bibinfo  {journal} {The Journal of Chemical Physics}\ }\textbf
  {\bibinfo {volume} {139}} (\bibinfo {year} {2013})}\BibitemShut {NoStop}%
\bibitem [{\citenamefont {Waheed}, \citenamefont {Ko},\ and\ \citenamefont
  {Rutledge}(2007)}]{Rutledge2007a}%
  \BibitemOpen
  \bibfield  {author} {\bibinfo {author} {\bibfnamefont {N.}~\bibnamefont
  {Waheed}}, \bibinfo {author} {\bibfnamefont {M.~J.}\ \bibnamefont {Ko}}, \
  and\ \bibinfo {author} {\bibfnamefont {G.~C.}\ \bibnamefont {Rutledge}},\
  }\bibfield  {title} {\enquote {\bibinfo {title} {Atomistic simulation of
  polymer melt crystallization by molecular dynamics},}\ }in\ \href@noop {}
  {\emph {\bibinfo {booktitle} {Progress in Understanding of Polymer
  Crystallization}}},\ \bibinfo {series and number} {Lect. Notes Phys. 714},\
  \bibinfo {editor} {edited by\ \bibinfo {editor} {\bibfnamefont
  {G.}~\bibnamefont {Reiter}}\ and\ \bibinfo {editor} {\bibfnamefont {G.~R.}\
  \bibnamefont {Strobl}}}\ (\bibinfo  {publisher} {Springer-Verlag},\ \bibinfo
  {address} {Berlin Heidelberg},\ \bibinfo {year} {2007})\ pp.\ \bibinfo
  {pages} {457--480}\BibitemShut {NoStop}%
\bibitem [{\citenamefont {Yi}, \citenamefont {Locker},\ and\ \citenamefont
  {Rutledge}(2013)}]{Yi2013}%
  \BibitemOpen
  \bibfield  {author} {\bibinfo {author} {\bibfnamefont {P.}~\bibnamefont
  {Yi}}, \bibinfo {author} {\bibfnamefont {C.~R.}\ \bibnamefont {Locker}}, \
  and\ \bibinfo {author} {\bibfnamefont {G.~C.}\ \bibnamefont {Rutledge}},\
  }\bibfield  {title} {\enquote {\bibinfo {title} {{Molecular dynamics
  simulation of homogeneous crystal nucleation in polyethylene}},}\ }\href
  {\doibase 10.1021/ma4004659} {\bibfield  {journal} {\bibinfo  {journal}
  {Macromolecules}\ }\textbf {\bibinfo {volume} {46}},\ \bibinfo {pages}
  {4723--4733} (\bibinfo {year} {2013})}\BibitemShut {NoStop}%
\bibitem [{\citenamefont {Yeh}, \citenamefont {Andzelm},\ and\ \citenamefont
  {Rutledge}(2015)}]{yeh2015mechanical}%
  \BibitemOpen
  \bibfield  {author} {\bibinfo {author} {\bibfnamefont {I.-C.}\ \bibnamefont
  {Yeh}}, \bibinfo {author} {\bibfnamefont {J.~W.}\ \bibnamefont {Andzelm}}, \
  and\ \bibinfo {author} {\bibfnamefont {G.~C.}\ \bibnamefont {Rutledge}},\
  }\bibfield  {title} {\enquote {\bibinfo {title} {Mechanical and structural
  characterization of semicrystalline polyethylene under tensile deformation by
  molecular dynamics simulations},}\ }\href@noop {} {\bibfield  {journal}
  {\bibinfo  {journal} {Macromolecules}\ }\textbf {\bibinfo {volume} {48}},\
  \bibinfo {pages} {4228--4239} (\bibinfo {year} {2015})}\BibitemShut {NoStop}%
\bibitem [{\citenamefont {Bourque}, \citenamefont {Locker},\ and\ \citenamefont
  {Rutledge}(2016)}]{bourque2016molecular}%
  \BibitemOpen
  \bibfield  {author} {\bibinfo {author} {\bibfnamefont {A.}~\bibnamefont
  {Bourque}}, \bibinfo {author} {\bibfnamefont {C.~R.}\ \bibnamefont {Locker}},
  \ and\ \bibinfo {author} {\bibfnamefont {G.~C.}\ \bibnamefont {Rutledge}},\
  }\bibfield  {title} {\enquote {\bibinfo {title} {Molecular dynamics
  simulation of surface nucleation during growth of an alkane crystal},}\
  }\href@noop {} {\bibfield  {journal} {\bibinfo  {journal} {Macromolecules}\
  }\textbf {\bibinfo {volume} {49}},\ \bibinfo {pages} {3619--3629} (\bibinfo
  {year} {2016})}\BibitemShut {NoStop}%
\bibitem [{\citenamefont {Sommer}(2007)}]{Sommer2007a}%
  \BibitemOpen
  \bibfield  {author} {\bibinfo {author} {\bibfnamefont {J.-U.}\ \bibnamefont
  {Sommer}},\ }\bibfield  {title} {\enquote {\bibinfo {title} {Theoretical
  aspects of the equilibrium state of chain crystals},}\ }in\ \href@noop {}
  {\emph {\bibinfo {booktitle} {Progress in Understanding of Polymer
  Crystallization}}},\ \bibinfo {series and number} {Lect. Notes Phys. 714},\
  \bibinfo {editor} {edited by\ \bibinfo {editor} {\bibfnamefont
  {G.}~\bibnamefont {Reiter}}\ and\ \bibinfo {editor} {\bibfnamefont {G.~R.}\
  \bibnamefont {Strobl}}}\ (\bibinfo  {publisher} {Springer-Verlag},\ \bibinfo
  {address} {Berlin Heidelberg},\ \bibinfo {year} {2007})\ pp.\ \bibinfo
  {pages} {19--45}\BibitemShut {NoStop}%
\bibitem [{\citenamefont {Luo}\ and\ \citenamefont {Sommer}(2014)}]{Luo2014}%
  \BibitemOpen
  \bibfield  {author} {\bibinfo {author} {\bibfnamefont {C.}~\bibnamefont
  {Luo}}\ and\ \bibinfo {author} {\bibfnamefont {J.~U.}\ \bibnamefont
  {Sommer}},\ }\bibfield  {title} {\enquote {\bibinfo {title} {{Frozen
  topology: Entanglements control nucleation and crystallization in
  polymers}},}\ }\href {\doibase 10.1103/PhysRevLett.112.195702} {\bibfield
  {journal} {\bibinfo  {journal} {Physical Review Letters}\ }\textbf {\bibinfo
  {volume} {112}},\ \bibinfo {pages} {195702} (\bibinfo {year}
  {2014})}\BibitemShut {NoStop}%
\bibitem [{\citenamefont {Luo}, \citenamefont {Kr\"oger},\ and\ \citenamefont
  {Sommer}(2016)}]{luo2016entanglements}%
  \BibitemOpen
  \bibfield  {author} {\bibinfo {author} {\bibfnamefont {C.}~\bibnamefont
  {Luo}}, \bibinfo {author} {\bibfnamefont {M.}~\bibnamefont {Kr\"oger}}, \
  and\ \bibinfo {author} {\bibfnamefont {J.-U.}\ \bibnamefont {Sommer}},\
  }\bibfield  {title} {\enquote {\bibinfo {title} {Entanglements and
  crystallization of concentrated polymer solutions: Molecular dynamics
  simulations},}\ }\href@noop {} {\bibfield  {journal} {\bibinfo  {journal}
  {Macromolecules}\ }\textbf {\bibinfo {volume} {49}},\ \bibinfo {pages}
  {9017--9025} (\bibinfo {year} {2016})}\BibitemShut {NoStop}%
\bibitem [{\citenamefont {Anwar}, \citenamefont {Turci},\ and\ \citenamefont
  {Schilling}(2013)}]{anwar2013crystallization}%
  \BibitemOpen
  \bibfield  {author} {\bibinfo {author} {\bibfnamefont {M.}~\bibnamefont
  {Anwar}}, \bibinfo {author} {\bibfnamefont {F.}~\bibnamefont {Turci}}, \ and\
  \bibinfo {author} {\bibfnamefont {T.}~\bibnamefont {Schilling}},\ }\bibfield
  {title} {\enquote {\bibinfo {title} {Crystallization mechanism in melts of
  short n-alkane chains},}\ }\href@noop {} {\bibfield  {journal} {\bibinfo
  {journal} {The Journal of Chemical Physics}\ }\textbf {\bibinfo {volume}
  {139}},\ \bibinfo {pages} {214904} (\bibinfo {year} {2013})}\BibitemShut
  {NoStop}%
\bibitem [{\citenamefont {Anwar}\ and\ \citenamefont
  {Schilling}(2015)}]{anwar2015crystallization}%
  \BibitemOpen
  \bibfield  {author} {\bibinfo {author} {\bibfnamefont {M.}~\bibnamefont
  {Anwar}}\ and\ \bibinfo {author} {\bibfnamefont {T.}~\bibnamefont
  {Schilling}},\ }\bibfield  {title} {\enquote {\bibinfo {title}
  {Crystallization of polyethylene: A molecular dynamics simulation study of
  the nucleation and growth mechanisms},}\ }\href@noop {} {\bibfield  {journal}
  {\bibinfo  {journal} {Polymer}\ }\textbf {\bibinfo {volume} {76}},\ \bibinfo
  {pages} {307--312} (\bibinfo {year} {2015})}\BibitemShut {NoStop}%
\bibitem [{\citenamefont {Wang}(2015)}]{Wang2015}%
  \BibitemOpen
  \bibfield  {author} {\bibinfo {author} {\bibfnamefont {M.-X.}\ \bibnamefont
  {Wang}},\ }\bibfield  {title} {\enquote {\bibinfo {title} {A single polymer
  folding and thickening from different dilute solution},}\ }\href {\doibase
  10.1016/j.physleta.2015.08.001} {\bibfield  {journal} {\bibinfo  {journal}
  {Physics Letters A}\ }\textbf {\bibinfo {volume} {379}},\ \bibinfo {pages}
  {2761--2765} (\bibinfo {year} {2015})}\BibitemShut {NoStop}%
\bibitem [{\citenamefont {Sadler}\ and\ \citenamefont
  {Keller}(1977)}]{Sadler77}%
  \BibitemOpen
  \bibfield  {author} {\bibinfo {author} {\bibfnamefont {D.~M.}\ \bibnamefont
  {Sadler}}\ and\ \bibinfo {author} {\bibfnamefont {A.}~\bibnamefont
  {Keller}},\ }\bibfield  {title} {\enquote {\bibinfo {title} {Neutron
  scattering studies on the molecular trajectory in polyethylene crystallized
  from solution and melt},}\ }\href {\doibase 10.1021/ma60059a045} {\bibfield
  {journal} {\bibinfo  {journal} {Macromolecules}\ }\textbf {\bibinfo {volume}
  {10}},\ \bibinfo {pages} {1128--1140} (\bibinfo {year} {1977})}\BibitemShut
  {NoStop}%
\bibitem [{\citenamefont {Fischer}\ \emph {et~al.}(1984)\citenamefont
  {Fischer}, \citenamefont {Hahn}, \citenamefont {Kugler}, \citenamefont
  {Struth}, \citenamefont {Born},\ and\ \citenamefont {Stamm}}]{Fischer1984}%
  \BibitemOpen
  \bibfield  {author} {\bibinfo {author} {\bibfnamefont {E.~W.}\ \bibnamefont
  {Fischer}}, \bibinfo {author} {\bibfnamefont {K.}~\bibnamefont {Hahn}},
  \bibinfo {author} {\bibfnamefont {J.}~\bibnamefont {Kugler}}, \bibinfo
  {author} {\bibfnamefont {U.}~\bibnamefont {Struth}}, \bibinfo {author}
  {\bibfnamefont {R.}~\bibnamefont {Born}}, \ and\ \bibinfo {author}
  {\bibfnamefont {M.}~\bibnamefont {Stamm}},\ }\bibfield  {title} {\enquote
  {\bibinfo {title} {An estimation of the number of tie molecules in
  semicrystalline polymers by means of neutron scattering},}\ }\href {\doibase
  10.1002/pol.1984.180220813} {\bibfield  {journal} {\bibinfo  {journal}
  {Journal of Polymer Science: Polymer Physics Edition}\ }\textbf {\bibinfo
  {volume} {22}},\ \bibinfo {pages} {1491--1513} (\bibinfo {year}
  {1984})}\BibitemShut {NoStop}%
\bibitem [{\citenamefont {Fischer}(1988)}]{Fischer1988}%
  \BibitemOpen
  \bibfield  {author} {\bibinfo {author} {\bibfnamefont {E.~W.}\ \bibnamefont
  {Fischer}},\ }\bibfield  {title} {\enquote {\bibinfo {title} {The
  conformation of polymer chains in the semicrystalline state},}\ }\href
  {\doibase 10.1002/masy.19880200130} {\bibfield  {journal} {\bibinfo
  {journal} {Makromolekulare Chemie, Macromolecular Symposia}\ }\textbf
  {\bibinfo {volume} {20/21}},\ \bibinfo {pages} {277--291} (\bibinfo {year}
  {1988})}\BibitemShut {NoStop}%
\bibitem [{\citenamefont {Hong}\ \emph {et~al.}(2015)\citenamefont {Hong},
  \citenamefont {Yuan}, \citenamefont {Li}, \citenamefont {Ke}, \citenamefont
  {Nozaki},\ and\ \citenamefont {Miyoshi}}]{HongPRL2015}%
  \BibitemOpen
  \bibfield  {author} {\bibinfo {author} {\bibfnamefont {Y.-l.}\ \bibnamefont
  {Hong}}, \bibinfo {author} {\bibfnamefont {S.}~\bibnamefont {Yuan}}, \bibinfo
  {author} {\bibfnamefont {Z.}~\bibnamefont {Li}}, \bibinfo {author}
  {\bibfnamefont {Y.}~\bibnamefont {Ke}}, \bibinfo {author} {\bibfnamefont
  {K.}~\bibnamefont {Nozaki}}, \ and\ \bibinfo {author} {\bibfnamefont
  {T.}~\bibnamefont {Miyoshi}},\ }\bibfield  {title} {\enquote {\bibinfo
  {title} {Three-dimensional conformation of folded polymers in single
  crystals},}\ }\href {\doibase 10.1103/PhysRevLett.115.168301} {\bibfield
  {journal} {\bibinfo  {journal} {Physical Review Letters}\ }\textbf {\bibinfo
  {volume} {115}},\ \bibinfo {pages} {168301} (\bibinfo {year}
  {2015})}\BibitemShut {NoStop}%
\bibitem [{\citenamefont {Hong}, \citenamefont {Koga},\ and\ \citenamefont
  {Miyoshi}(2015)}]{HongMacromol2015}%
  \BibitemOpen
  \bibfield  {author} {\bibinfo {author} {\bibfnamefont {Y.-l.}\ \bibnamefont
  {Hong}}, \bibinfo {author} {\bibfnamefont {T.}~\bibnamefont {Koga}}, \ and\
  \bibinfo {author} {\bibfnamefont {T.}~\bibnamefont {Miyoshi}},\ }\bibfield
  {title} {\enquote {\bibinfo {title} {Chain trajectory and crystallization
  mechanism of a semicrystalline polymer in melt-- and solution--grown crystals
  as studied using ${}^{13}${C}--${}^{13}${C} double-quantum {NMR}},}\ }\href
  {\doibase 10.1021/acs.macromol.5b00079} {\bibfield  {journal} {\bibinfo
  {journal} {Macromolecules}\ }\textbf {\bibinfo {volume} {48}},\ \bibinfo
  {pages} {3282--3293} (\bibinfo {year} {2015})}\BibitemShut {NoStop}%
\bibitem [{\citenamefont {Hong}\ \emph {et~al.}(2016)\citenamefont {Hong},
  \citenamefont {Chen}, \citenamefont {Yuan}, \citenamefont {Kang},\ and\
  \citenamefont {Miyoshi}}]{HongACSLett2016}%
  \BibitemOpen
  \bibfield  {author} {\bibinfo {author} {\bibfnamefont {Y.-l.}\ \bibnamefont
  {Hong}}, \bibinfo {author} {\bibfnamefont {W.}~\bibnamefont {Chen}}, \bibinfo
  {author} {\bibfnamefont {S.}~\bibnamefont {Yuan}}, \bibinfo {author}
  {\bibfnamefont {J.}~\bibnamefont {Kang}}, \ and\ \bibinfo {author}
  {\bibfnamefont {T.}~\bibnamefont {Miyoshi}},\ }\bibfield  {title} {\enquote
  {\bibinfo {title} {Chain trajectory of semicrystalline polymers as revealed
  by solid--state {NMR} spectroscopy},}\ }\href {\doibase
  10.1021/acsmacrolett.6b00040} {\bibfield  {journal} {\bibinfo  {journal} {ACS
  Macro Letters}\ }\textbf {\bibinfo {volume} {5}},\ \bibinfo {pages}
  {355--358} (\bibinfo {year} {2016})}\BibitemShut {NoStop}%
\bibitem [{\citenamefont {Grosberg}, \citenamefont {Nechaev},\ and\
  \citenamefont {Shakhnovich}(1988)}]{Nechaev}%
  \BibitemOpen
  \bibfield  {author} {\bibinfo {author} {\bibfnamefont {A.~Y.}\ \bibnamefont
  {Grosberg}}, \bibinfo {author} {\bibfnamefont {S.~K.}\ \bibnamefont
  {Nechaev}}, \ and\ \bibinfo {author} {\bibfnamefont {E.~I.}\ \bibnamefont
  {Shakhnovich}},\ }\bibfield  {title} {\enquote {\bibinfo {title} {The role of
  topological constraints in the kinetics of collapse of macromolecules},}\
  }\href {\doibase 10.1051/jphys:0198800490120209500} {\bibfield  {journal}
  {\bibinfo  {journal} {Journal de physique France}\ }\textbf {\bibinfo
  {volume} {49}},\ \bibinfo {pages} {2095--2100} (\bibinfo {year}
  {1988})}\BibitemShut {NoStop}%
\bibitem [{\citenamefont {Chertovich}\ and\ \citenamefont
  {Kos}(2014)}]{chertovich2014crumpled}%
  \BibitemOpen
  \bibfield  {author} {\bibinfo {author} {\bibfnamefont {A.}~\bibnamefont
  {Chertovich}}\ and\ \bibinfo {author} {\bibfnamefont {P.}~\bibnamefont
  {Kos}},\ }\bibfield  {title} {\enquote {\bibinfo {title} {Crumpled globule
  formation during collapse of a long flexible and semiflexible polymer in poor
  solvent},}\ }\href@noop {} {\bibfield  {journal} {\bibinfo  {journal} {The
  Journal of Chemical Physics}\ }\textbf {\bibinfo {volume} {141}},\ \bibinfo
  {pages} {134903} (\bibinfo {year} {2014})}\BibitemShut {NoStop}%
\bibitem [{\citenamefont {Hu}, \citenamefont {Frenkel},\ and\ \citenamefont
  {Mathot}(2003)}]{HuFrenkel}%
  \BibitemOpen
  \bibfield  {author} {\bibinfo {author} {\bibfnamefont {W.}~\bibnamefont
  {Hu}}, \bibinfo {author} {\bibfnamefont {D.}~\bibnamefont {Frenkel}}, \ and\
  \bibinfo {author} {\bibfnamefont {V.~B.~F.}\ \bibnamefont {Mathot}},\
  }\bibfield  {title} {\enquote {\bibinfo {title} {Intramolecular nucleation
  model for polymer crystallization},}\ }\href {\doibase 10.1021/ma0344285}
  {\bibfield  {journal} {\bibinfo  {journal} {Macromolecules}\ }\textbf
  {\bibinfo {volume} {36}},\ \bibinfo {pages} {8178--8183} (\bibinfo {year}
  {2003})}\BibitemShut {NoStop}%
\bibitem [{\citenamefont {Hu}\ and\ \citenamefont
  {Frenkel}(2006)}]{HuFrenkel2}%
  \BibitemOpen
  \bibfield  {author} {\bibinfo {author} {\bibfnamefont {W.}~\bibnamefont
  {Hu}}\ and\ \bibinfo {author} {\bibfnamefont {D.}~\bibnamefont {Frenkel}},\
  }\bibfield  {title} {\enquote {\bibinfo {title} {Effect of the coil-globule
  transition on the free-energy barrier for intrachain crystal nucleation},}\
  }\href@noop {} {\bibfield  {journal} {\bibinfo  {journal} {J. Phys. Chem. B}\
  }\textbf {\bibinfo {volume} {110}},\ \bibinfo {pages} {3734--3737} (\bibinfo
  {year} {2006})}\BibitemShut {NoStop}%
\bibitem [{\citenamefont {Hu}\ and\ \citenamefont {Cai}(2008)}]{Hu3}%
  \BibitemOpen
  \bibfield  {author} {\bibinfo {author} {\bibfnamefont {W.}~\bibnamefont
  {Hu}}\ and\ \bibinfo {author} {\bibfnamefont {T.}~\bibnamefont {Cai}},\
  }\bibfield  {title} {\enquote {\bibinfo {title} {Regime transitions of
  polymer crystal growth rates: Molecular simulations and interpretation beyond
  lauritzen-hoffman model},}\ }\href@noop {} {\bibfield  {journal} {\bibinfo
  {journal} {Macromolecules}\ }\textbf {\bibinfo {volume} {41}},\ \bibinfo
  {pages} {2049--2061} (\bibinfo {year} {2008})}\BibitemShut {NoStop}%
\bibitem [{\citenamefont {Onsager}(1949)}]{Onsager}%
  \BibitemOpen
  \bibfield  {author} {\bibinfo {author} {\bibfnamefont {L.}~\bibnamefont
  {Onsager}},\ }\bibfield  {title} {\enquote {\bibinfo {title} {The effects of
  shape on the interaction of colloidal particles},}\ }\href {\doibase
  10.1111/j.1749-6632.1949.tb27296.x} {\bibfield  {journal} {\bibinfo
  {journal} {Annals of the New York Academy of Sciences}\ }\textbf {\bibinfo
  {volume} {51}},\ \bibinfo {pages} {627--659} (\bibinfo {year}
  {1949})}\BibitemShut {NoStop}%
\bibitem [{\citenamefont {Khokhlov}\ and\ \citenamefont
  {Semenov}(1985)}]{KhoSem}%
  \BibitemOpen
  \bibfield  {author} {\bibinfo {author} {\bibfnamefont {A.}~\bibnamefont
  {Khokhlov}}\ and\ \bibinfo {author} {\bibfnamefont {A.}~\bibnamefont
  {Semenov}},\ }\bibfield  {title} {\enquote {\bibinfo {title} {On the theory
  of liquid-crystalline ordering of polymer chains with limited flexibility},}\
  }\href {\doibase 10.1007/BF01017855} {\bibfield  {journal} {\bibinfo
  {journal} {J. Stat. Phys.}\ }\textbf {\bibinfo {volume} {38}},\ \bibinfo
  {pages} {161--182} (\bibinfo {year} {1985})}\BibitemShut {NoStop}%
\bibitem [{\citenamefont {Groot}\ and\ \citenamefont
  {Warren}(1997)}]{groot1997dissipative}%
  \BibitemOpen
  \bibfield  {author} {\bibinfo {author} {\bibfnamefont {R.~D.}\ \bibnamefont
  {Groot}}\ and\ \bibinfo {author} {\bibfnamefont {P.~B.}\ \bibnamefont
  {Warren}},\ }\bibfield  {title} {\enquote {\bibinfo {title} {Dissipative
  particle dynamics: Bridging the gap between atomistic and mesoscopic
  simulation},}\ }\href@noop {} {\bibfield  {journal} {\bibinfo  {journal} {The
  Journal of Chemical Physics}\ }\textbf {\bibinfo {volume} {107}},\ \bibinfo
  {pages} {4423--4435} (\bibinfo {year} {1997})}\BibitemShut {NoStop}%
\bibitem [{\citenamefont {Espa{\~n}ol}\ and\ \citenamefont
  {Warren}(2017)}]{espanol2017perspective}%
  \BibitemOpen
  \bibfield  {author} {\bibinfo {author} {\bibfnamefont {P.}~\bibnamefont
  {Espa{\~n}ol}}\ and\ \bibinfo {author} {\bibfnamefont {P.~B.}\ \bibnamefont
  {Warren}},\ }\bibfield  {title} {\enquote {\bibinfo {title} {Perspective:
  Dissipative particle dynamics},}\ }\href@noop {} {\bibfield  {journal}
  {\bibinfo  {journal} {The Journal of Chemical Physics}\ }\textbf {\bibinfo
  {volume} {146}},\ \bibinfo {pages} {150901} (\bibinfo {year}
  {2017})}\BibitemShut {NoStop}%
\bibitem [{\citenamefont {Gavrilov}\ \emph {et~al.}(2011)\citenamefont
  {Gavrilov}, \citenamefont {Kudryavtsev}, \citenamefont {Khalatur},\ and\
  \citenamefont {Chertovich}}]{gavrilov2011phase}%
  \BibitemOpen
  \bibfield  {author} {\bibinfo {author} {\bibfnamefont {A.}~\bibnamefont
  {Gavrilov}}, \bibinfo {author} {\bibfnamefont {Y.}~\bibnamefont
  {Kudryavtsev}}, \bibinfo {author} {\bibfnamefont {P.}~\bibnamefont
  {Khalatur}}, \ and\ \bibinfo {author} {\bibfnamefont {A.}~\bibnamefont
  {Chertovich}},\ }\bibfield  {title} {\enquote {\bibinfo {title} {Microphase
  separation in regular and random copolymer melts by {DPD} simulations},}\
  }\href@noop {} {\bibfield  {journal} {\bibinfo  {journal} {Chem. Phys.
  Lett.}\ }\textbf {\bibinfo {volume} {503}},\ \bibinfo {pages} {277--282}
  (\bibinfo {year} {2011})}\BibitemShut {NoStop}%
\bibitem [{\citenamefont {Gavrilov}, \citenamefont {Kudryavtsev},\ and\
  \citenamefont {Chertovich}(2013)}]{gavrilov2013phase}%
  \BibitemOpen
  \bibfield  {author} {\bibinfo {author} {\bibfnamefont {A.~A.}\ \bibnamefont
  {Gavrilov}}, \bibinfo {author} {\bibfnamefont {Y.~V.}\ \bibnamefont
  {Kudryavtsev}}, \ and\ \bibinfo {author} {\bibfnamefont {A.~V.}\ \bibnamefont
  {Chertovich}},\ }\bibfield  {title} {\enquote {\bibinfo {title} {Phase
  diagrams of block copolymer melts by dissipative particle dynamics
  simulations},}\ }\href@noop {} {\bibfield  {journal} {\bibinfo  {journal}
  {The Journal of Chemical Physics}\ }\textbf {\bibinfo {volume} {139}},\
  \bibinfo {pages} {224901} (\bibinfo {year} {2013})}\BibitemShut {NoStop}%
\bibitem [{\citenamefont {Nikunen}, \citenamefont {Vattulainen},\ and\
  \citenamefont {Karttunen}(2007)}]{nikunen2007reptational}%
  \BibitemOpen
  \bibfield  {author} {\bibinfo {author} {\bibfnamefont {P.}~\bibnamefont
  {Nikunen}}, \bibinfo {author} {\bibfnamefont {I.}~\bibnamefont
  {Vattulainen}}, \ and\ \bibinfo {author} {\bibfnamefont {M.}~\bibnamefont
  {Karttunen}},\ }\bibfield  {title} {\enquote {\bibinfo {title} {Reptational
  dynamics in dissipative particle dynamics simulations of polymer melts},}\
  }\href@noop {} {\bibfield  {journal} {\bibinfo  {journal} {Physical Review
  E}\ }\textbf {\bibinfo {volume} {75}},\ \bibinfo {pages} {036713} (\bibinfo
  {year} {2007})}\BibitemShut {NoStop}%
\bibitem [{\citenamefont {Vega}, \citenamefont {McBride},\ and\ \citenamefont
  {MacDowell}(2001)}]{vega2001liquid}%
  \BibitemOpen
  \bibfield  {author} {\bibinfo {author} {\bibfnamefont {C.}~\bibnamefont
  {Vega}}, \bibinfo {author} {\bibfnamefont {C.}~\bibnamefont {McBride}}, \
  and\ \bibinfo {author} {\bibfnamefont {L.~G.}\ \bibnamefont {MacDowell}},\
  }\bibfield  {title} {\enquote {\bibinfo {title} {Liquid crystal phase
  formation for the linear tangent hard sphere model from monte carlo
  simulations},}\ }\href@noop {} {\bibfield  {journal} {\bibinfo  {journal}
  {The Journal of Chemical Physics}\ }\textbf {\bibinfo {volume} {115}},\
  \bibinfo {pages} {4203--4211} (\bibinfo {year} {2001})}\BibitemShut {NoStop}%
\bibitem [{\citenamefont {Raos}\ and\ \citenamefont
  {Allegra}(1997)}]{Raos1997}%
  \BibitemOpen
  \bibfield  {author} {\bibinfo {author} {\bibfnamefont {G.}~\bibnamefont
  {Raos}}\ and\ \bibinfo {author} {\bibfnamefont {G.}~\bibnamefont {Allegra}},\
  }\bibfield  {title} {\enquote {\bibinfo {title} {Macromolecular clusters in
  poor-solvent polymer solutions},}\ }\href {\doibase 10.1063/1.474306}
  {\bibfield  {journal} {\bibinfo  {journal} {The Journal of Chemical Physics}\
  }\textbf {\bibinfo {volume} {107}},\ \bibinfo {pages} {6479--6490} (\bibinfo
  {year} {1997})}\BibitemShut {NoStop}%
\bibitem [{\citenamefont {Eichhorn}\ \emph {et~al.}(2009)\citenamefont
  {Eichhorn}, \citenamefont {Hearle}, \citenamefont {Jaffe},\ and\
  \citenamefont {Kikutani}}]{eichhorn2009handbook}%
  \BibitemOpen
  \bibfield  {author} {\bibinfo {author} {\bibfnamefont {S.}~\bibnamefont
  {Eichhorn}}, \bibinfo {author} {\bibfnamefont {J.}~\bibnamefont {Hearle}},
  \bibinfo {author} {\bibfnamefont {M.}~\bibnamefont {Jaffe}}, \ and\ \bibinfo
  {author} {\bibfnamefont {T.}~\bibnamefont {Kikutani}},\ }\href@noop {} {\emph
  {\bibinfo {title} {Handbook of Textile Fibre Structure: Volume 2: Natural,
  Regenerated, inorganic and Specialist Fibres}}}\ (\bibinfo  {publisher}
  {Elsevier},\ \bibinfo {year} {2009})\BibitemShut {NoStop}%
\bibitem [{\citenamefont {Spenley}(2000)}]{spenley2000scaling}%
  \BibitemOpen
  \bibfield  {author} {\bibinfo {author} {\bibfnamefont {N.}~\bibnamefont
  {Spenley}},\ }\bibfield  {title} {\enquote {\bibinfo {title} {Scaling laws
  for polymers in dissipative particle dynamics},}\ }\href@noop {} {\bibfield
  {journal} {\bibinfo  {journal} {Europhysics Letters}\ }\textbf {\bibinfo
  {volume} {49}},\ \bibinfo {pages} {534} (\bibinfo {year} {2000})}\BibitemShut
  {NoStop}%
\bibitem [{\citenamefont {Lahmar}\ and\ \citenamefont
  {Rousseau}(2007)}]{lahmar2007influence}%
  \BibitemOpen
  \bibfield  {author} {\bibinfo {author} {\bibfnamefont {F.}~\bibnamefont
  {Lahmar}}\ and\ \bibinfo {author} {\bibfnamefont {B.}~\bibnamefont
  {Rousseau}},\ }\bibfield  {title} {\enquote {\bibinfo {title} {Influence of
  the adjustable parameters of the {DPD} on the global and local dynamics of a
  polymer melt},}\ }\href@noop {} {\bibfield  {journal} {\bibinfo  {journal}
  {Polymer}\ }\textbf {\bibinfo {volume} {48}},\ \bibinfo {pages} {3584--3592}
  (\bibinfo {year} {2007})}\BibitemShut {NoStop}%
\bibitem [{\citenamefont {Karatrantos}\ \emph {et~al.}(2013)\citenamefont
  {Karatrantos}, \citenamefont {Clarke}, \citenamefont {Composto},\ and\
  \citenamefont {Winey}}]{Karat-Winey-2013}%
  \BibitemOpen
  \bibfield  {author} {\bibinfo {author} {\bibfnamefont {A.}~\bibnamefont
  {Karatrantos}}, \bibinfo {author} {\bibfnamefont {N.}~\bibnamefont {Clarke}},
  \bibinfo {author} {\bibfnamefont {R.~J.}\ \bibnamefont {Composto}}, \ and\
  \bibinfo {author} {\bibfnamefont {K.~I.}\ \bibnamefont {Winey}},\ }\bibfield
  {title} {\enquote {\bibinfo {title} {Topological entanglement length in
  polymer melts and nanocomposites by a {DPD} polymer model},}\ }\href
  {\doibase 10.1039/c3sm27651a} {\bibfield  {journal} {\bibinfo  {journal}
  {Soft Matter}\ }\textbf {\bibinfo {volume} {9}},\ \bibinfo {pages}
  {3877--3884} (\bibinfo {year} {2013})}\BibitemShut {NoStop}%
\bibitem [{\citenamefont {Chang}\ and\ \citenamefont
  {Yethiraj}(2019)}]{chang2019can}%
  \BibitemOpen
  \bibfield  {author} {\bibinfo {author} {\bibfnamefont {R.}~\bibnamefont
  {Chang}}\ and\ \bibinfo {author} {\bibfnamefont {A.}~\bibnamefont
  {Yethiraj}},\ }\bibfield  {title} {\enquote {\bibinfo {title} {Can polymer
  chains cross each other and still be entangled?}}\ }\href@noop {} {\bibfield
  {journal} {\bibinfo  {journal} {Macromolecules}\ }\textbf {\bibinfo {volume}
  {52}},\ \bibinfo {pages} {2000--2006} (\bibinfo {year} {2019})}\BibitemShut
  {NoStop}%
\bibitem [{\citenamefont {Markina}\ \emph {et~al.}(2017)\citenamefont
  {Markina}, \citenamefont {Ivanov}, \citenamefont {Komarov}, \citenamefont
  {Larin}, \citenamefont {Kenny},\ and\ \citenamefont {Lyulin}}]{MarkinaPI}%
  \BibitemOpen
  \bibfield  {author} {\bibinfo {author} {\bibfnamefont {A.}~\bibnamefont
  {Markina}}, \bibinfo {author} {\bibfnamefont {V.}~\bibnamefont {Ivanov}},
  \bibinfo {author} {\bibfnamefont {P.}~\bibnamefont {Komarov}}, \bibinfo
  {author} {\bibfnamefont {S.}~\bibnamefont {Larin}}, \bibinfo {author}
  {\bibfnamefont {J.~M.}\ \bibnamefont {Kenny}}, \ and\ \bibinfo {author}
  {\bibfnamefont {S.}~\bibnamefont {Lyulin}},\ }\bibfield  {title} {\enquote
  {\bibinfo {title} {Effect of polymer chain stiffness on initial stages of
  crystallization of polyetherimides: Coarse-grained computer simulation},}\
  }\href {\doibase 10.1002/polb.24380} {\bibfield  {journal} {\bibinfo
  {journal} {Journal of Polymer Science, Part B: Polymer Physics}\ }\textbf
  {\bibinfo {volume} {55}},\ \bibinfo {pages} {1254--1265} (\bibinfo {year}
  {2017})}\BibitemShut {NoStop}%
\bibitem [{\citenamefont {Gowers}\ \emph {et~al.}(2016)\citenamefont {Gowers},
  \citenamefont {Linke}, \citenamefont {Barnoud}, \citenamefont {Reddy},
  \citenamefont {Melo}, \citenamefont {Seyler}, \citenamefont {Dotson},
  \citenamefont {Domanski}, \citenamefont {Buchoux},\ and\ \citenamefont
  {Kenney}}]{gowers2016mdanalysis}%
  \BibitemOpen
  \bibfield  {author} {\bibinfo {author} {\bibfnamefont {R.~J.}\ \bibnamefont
  {Gowers}}, \bibinfo {author} {\bibfnamefont {M.}~\bibnamefont {Linke}},
  \bibinfo {author} {\bibfnamefont {J.}~\bibnamefont {Barnoud}}, \bibinfo
  {author} {\bibfnamefont {T.~J.}\ \bibnamefont {Reddy}}, \bibinfo {author}
  {\bibfnamefont {M.~N.}\ \bibnamefont {Melo}}, \bibinfo {author}
  {\bibfnamefont {S.~L.}\ \bibnamefont {Seyler}}, \bibinfo {author}
  {\bibfnamefont {D.~L.}\ \bibnamefont {Dotson}}, \bibinfo {author}
  {\bibfnamefont {J.}~\bibnamefont {Domanski}}, \bibinfo {author}
  {\bibfnamefont {S.}~\bibnamefont {Buchoux}}, \ and\ \bibinfo {author}
  {\bibfnamefont {I.~M.}\ \bibnamefont {Kenney}},\ }\bibfield  {title}
  {\enquote {\bibinfo {title} {Mdanalysis: a python package for the rapid
  analysis of molecular dynamics simulations},}\ }in\ \href@noop {} {\emph
  {\bibinfo {booktitle} {Proceedings of the 15th Python in Science
  Conference}}},\ \bibinfo {editor} {edited by\ \bibinfo {editor}
  {\bibfnamefont {S.}~\bibnamefont {Benthall}}\ and\ \bibinfo {editor}
  {\bibfnamefont {S.}~\bibnamefont {Rostrup}}}\ (\bibinfo  {publisher}
  {Almquist \& Wiksell},\ \bibinfo {year} {2016})\ pp.\ \bibinfo {pages}
  {98--105}\BibitemShut {NoStop}%
\bibitem [{\citenamefont {Michaud-Agrawal}\ \emph {et~al.}(2011)\citenamefont
  {Michaud-Agrawal}, \citenamefont {Denning}, \citenamefont {Woolf},\ and\
  \citenamefont {Beckstein}}]{michaud2011mdanalysis}%
  \BibitemOpen
  \bibfield  {author} {\bibinfo {author} {\bibfnamefont {N.}~\bibnamefont
  {Michaud-Agrawal}}, \bibinfo {author} {\bibfnamefont {E.~J.}\ \bibnamefont
  {Denning}}, \bibinfo {author} {\bibfnamefont {T.~B.}\ \bibnamefont {Woolf}},
  \ and\ \bibinfo {author} {\bibfnamefont {O.}~\bibnamefont {Beckstein}},\
  }\bibfield  {title} {\enquote {\bibinfo {title} {Mdanalysis: a toolkit for
  the analysis of molecular dynamics simulations},}\ }\href@noop {} {\bibfield
  {journal} {\bibinfo  {journal} {Journal of Computational Chemistry}\ }\textbf
  {\bibinfo {volume} {32}},\ \bibinfo {pages} {2319--2327} (\bibinfo {year}
  {2011})}\BibitemShut {NoStop}%
\bibitem [{\citenamefont {Ivanov}\ \emph {et~al.}(2014)\citenamefont {Ivanov},
  \citenamefont {Rodionova}, \citenamefont {Martemyanova}, \citenamefont
  {Stukan}, \citenamefont {M\"uller}, \citenamefont {Paul},\ and\ \citenamefont
  {Binder}}]{Ivanov2014}%
  \BibitemOpen
  \bibfield  {author} {\bibinfo {author} {\bibfnamefont {V.~A.}\ \bibnamefont
  {Ivanov}}, \bibinfo {author} {\bibfnamefont {A.~S.}\ \bibnamefont
  {Rodionova}}, \bibinfo {author} {\bibfnamefont {J.~A.}\ \bibnamefont
  {Martemyanova}}, \bibinfo {author} {\bibfnamefont {M.~R.}\ \bibnamefont
  {Stukan}}, \bibinfo {author} {\bibfnamefont {M.}~\bibnamefont {M\"uller}},
  \bibinfo {author} {\bibfnamefont {W.}~\bibnamefont {Paul}}, \ and\ \bibinfo
  {author} {\bibfnamefont {K.}~\bibnamefont {Binder}},\ }\bibfield  {title}
  {\enquote {\bibinfo {title} {Conformational properties of semiflexible chains
  at nematic ordering transitions in thin films: a monte carlo simulation},}\
  }\href@noop {} {\bibfield  {journal} {\bibinfo  {journal} {Macromolecules}\
  }\textbf {\bibinfo {volume} {47}},\ \bibinfo {pages} {1206--1220} (\bibinfo
  {year} {2014})}\BibitemShut {NoStop}%
\bibitem [{\citenamefont {Allen}\ and\ \citenamefont
  {Tildesley}(1989)}]{allen1989computer}%
  \BibitemOpen
  \bibfield  {author} {\bibinfo {author} {\bibfnamefont {M.~P.}\ \bibnamefont
  {Allen}}\ and\ \bibinfo {author} {\bibfnamefont {D.~J.}\ \bibnamefont
  {Tildesley}},\ }\href@noop {} {\emph {\bibinfo {title} {Computer Simulation
  of Liquids}}}\ (\bibinfo  {publisher} {Oxford University Press},\ \bibinfo
  {year} {1989})\BibitemShut {NoStop}%
\bibitem [{\citenamefont {Lifshitz}, \citenamefont {Grosberg},\ and\
  \citenamefont {Khokhlov}(1978)}]{lifshitz1978some}%
  \BibitemOpen
  \bibfield  {author} {\bibinfo {author} {\bibfnamefont {I.}~\bibnamefont
  {Lifshitz}}, \bibinfo {author} {\bibfnamefont {A.~Y.}\ \bibnamefont
  {Grosberg}}, \ and\ \bibinfo {author} {\bibfnamefont {A.}~\bibnamefont
  {Khokhlov}},\ }\bibfield  {title} {\enquote {\bibinfo {title} {Some problems
  of the statistical physics of polymer chains with volume interaction},}\
  }\href@noop {} {\bibfield  {journal} {\bibinfo  {journal} {Reviews of Modern
  Physics}\ }\textbf {\bibinfo {volume} {50}},\ \bibinfo {pages} {683}
  (\bibinfo {year} {1978})}\BibitemShut {NoStop}%
\bibitem [{\citenamefont {Kr{\"o}ger}(2005)}]{kroger2005shortest}%
  \BibitemOpen
  \bibfield  {author} {\bibinfo {author} {\bibfnamefont {M.}~\bibnamefont
  {Kr{\"o}ger}},\ }\bibfield  {title} {\enquote {\bibinfo {title} {Shortest
  multiple disconnected path for the analysis of entanglements in two-and
  three-dimensional polymeric systems},}\ }\href@noop {} {\bibfield  {journal}
  {\bibinfo  {journal} {Computer Physics Communications}\ }\textbf {\bibinfo
  {volume} {168}},\ \bibinfo {pages} {209--232} (\bibinfo {year}
  {2005})}\BibitemShut {NoStop}%
\bibitem [{\citenamefont {Shanbhag}\ and\ \citenamefont
  {Kr{\"o}ger}(2007)}]{shanbhag2007primitive}%
  \BibitemOpen
  \bibfield  {author} {\bibinfo {author} {\bibfnamefont {S.}~\bibnamefont
  {Shanbhag}}\ and\ \bibinfo {author} {\bibfnamefont {M.}~\bibnamefont
  {Kr{\"o}ger}},\ }\bibfield  {title} {\enquote {\bibinfo {title} {Primitive
  path networks generated by annealing and geometrical methods: Insights into
  differences},}\ }\href@noop {} {\bibfield  {journal} {\bibinfo  {journal}
  {Macromolecules}\ }\textbf {\bibinfo {volume} {40}},\ \bibinfo {pages}
  {2897--2903} (\bibinfo {year} {2007})}\BibitemShut {NoStop}%
\bibitem [{\citenamefont {Hoy}, \citenamefont {Foteinopoulou},\ and\
  \citenamefont {Kr{\"o}ger}(2009)}]{hoy2009topological}%
  \BibitemOpen
  \bibfield  {author} {\bibinfo {author} {\bibfnamefont {R.~S.}\ \bibnamefont
  {Hoy}}, \bibinfo {author} {\bibfnamefont {K.}~\bibnamefont {Foteinopoulou}},
  \ and\ \bibinfo {author} {\bibfnamefont {M.}~\bibnamefont {Kr{\"o}ger}},\
  }\bibfield  {title} {\enquote {\bibinfo {title} {Topological analysis of
  polymeric melts: Chain-length effects and fast-converging estimators for
  entanglement length},}\ }\href@noop {} {\bibfield  {journal} {\bibinfo
  {journal} {Physical Review E}\ }\textbf {\bibinfo {volume} {80}},\ \bibinfo
  {pages} {031803} (\bibinfo {year} {2009})}\BibitemShut {NoStop}%
\bibitem [{\citenamefont {Karayiannis}\ and\ \citenamefont
  {Kr{\"o}ger}(2009)}]{karayiannis2009combined}%
  \BibitemOpen
  \bibfield  {author} {\bibinfo {author} {\bibfnamefont {N.~C.}\ \bibnamefont
  {Karayiannis}}\ and\ \bibinfo {author} {\bibfnamefont {M.}~\bibnamefont
  {Kr{\"o}ger}},\ }\bibfield  {title} {\enquote {\bibinfo {title} {Combined
  molecular algorithms for the generation, equilibration and topological
  analysis of entangled polymers: Methodology and performance},}\ }\href@noop
  {} {\bibfield  {journal} {\bibinfo  {journal} {International Journal of
  Molecular Sciences}\ }\textbf {\bibinfo {volume} {10}},\ \bibinfo {pages}
  {5054--5089} (\bibinfo {year} {2009})}\BibitemShut {NoStop}%
\bibitem [{\citenamefont {Khokhlov}\ and\ \citenamefont
  {Nechaev}(1985)}]{KhokhlovNechaev}%
  \BibitemOpen
  \bibfield  {author} {\bibinfo {author} {\bibfnamefont {A.~R.}\ \bibnamefont
  {Khokhlov}}\ and\ \bibinfo {author} {\bibfnamefont {S.~K.}\ \bibnamefont
  {Nechaev}},\ }\bibfield  {title} {\enquote {\bibinfo {title} {Polymer chain
  in an array of obstacles},}\ }\href@noop {} {\bibfield  {journal} {\bibinfo
  {journal} {Physics Letters A}\ }\textbf {\bibinfo {volume} {112}},\ \bibinfo
  {pages} {156--160} (\bibinfo {year} {1985})}\BibitemShut {NoStop}%
\bibitem [{\citenamefont {Voevodin}\ \emph {et~al.}(2019)\citenamefont
  {Voevodin}, \citenamefont {Antonov}, \citenamefont {Nikitenko}, \citenamefont
  {Shvets}, \citenamefont {Sobolev}, \citenamefont {Stefanov}, \citenamefont
  {Voevodin}, \citenamefont {Zhumatiy}, \citenamefont {Brechalov},\ and\
  \citenamefont {Naumov}}]{voevodin2019lomonosov}%
  \BibitemOpen
  \bibfield  {author} {\bibinfo {author} {\bibfnamefont {V.}~\bibnamefont
  {Voevodin}}, \bibinfo {author} {\bibfnamefont {A.}~\bibnamefont {Antonov}},
  \bibinfo {author} {\bibfnamefont {D.}~\bibnamefont {Nikitenko}}, \bibinfo
  {author} {\bibfnamefont {P.}~\bibnamefont {Shvets}}, \bibinfo {author}
  {\bibfnamefont {S.}~\bibnamefont {Sobolev}}, \bibinfo {author} {\bibfnamefont
  {K.}~\bibnamefont {Stefanov}}, \bibinfo {author} {\bibfnamefont
  {V.}~\bibnamefont {Voevodin}}, \bibinfo {author} {\bibfnamefont
  {S.}~\bibnamefont {Zhumatiy}}, \bibinfo {author} {\bibfnamefont
  {A.}~\bibnamefont {Brechalov}}, \ and\ \bibinfo {author} {\bibfnamefont
  {A.}~\bibnamefont {Naumov}},\ }\bibfield  {title} {\enquote {\bibinfo {title}
  {Lomonosov-2: Petascale supercomputing at lomonosov moscow state
  university},}\ }in\ \href@noop {} {\emph {\bibinfo {booktitle} {Contemporary
  High Performance Computing: From Petascale toward Exascale}}},\ \bibinfo
  {series} {Chapman \& Hall/CRC Computational Science Series}, Vol.~\bibinfo
  {volume} {3},\ \bibinfo {editor} {edited by\ \bibinfo {editor} {\bibfnamefont
  {J.~S.}\ \bibnamefont {Vetter}}}\ (\bibinfo  {publisher} {CRC Press},\
  \bibinfo {year} {2019})\ pp.\ \bibinfo {pages} {305--330}\BibitemShut
  {NoStop}%
\end{thebibliography}%

\end{document}